\documentclass[us eAMS,usenatbib]{mn2e}
\usepackage{times}
\usepackage{graphicx}
\usepackage{aas_macros}
\usepackage[usenames]{color}
\definecolor{Comm}{rgb}{0.2,0.2,0.8}
\definecolor{Todo}{rgb}{0.7,0,0}
\definecolor{Old}{rgb}{0.2,0.7,0.2}
\title[{\it Kepler} observations of variability in B-type stars]
{{\it Kepler} observations of variability in B-type stars}
\author[L.A. Balona et. al.]
{L. A. Balona$^{1}$, A. Pigulski$^{2}$, P. De Cat$^{3}$, G. Handler$^{4}$, J.
Guti\'errez-Soto$^{5,15}$,
\newauthor{C. A. Engelbrecht$^{6}$, F. Frescura$^{7}$, M. Briquet$^{8}$, J. Cuypers$^{3}$, J. Daszy\'nska-Daszkiewicz$^{2}$,}
\newauthor{P. Degroote$^{8}$, R. J. Dukes$^{9}$, R. A. Garcia$^{10}$, E. M. Green$^{11}$, U. Heber$^{12}$, S. D. Kawaler$^{13}$,}
\newauthor{H. Lehmann$^{14}$, B. Leroy$^{15}$, J. Molenda-\.Zakowicz$^{2}$, C. Neiner$^{15}$, A. Noels$^{16}$, J. Nuspl$^{17}$,}
\newauthor{R. {\O}stensen$^{8}$, D. Pricopi$^{18}$, I. Roxburgh$^{17}$, S. Salmon$^{16}$, M. A. Smith$^{19}$, J. C. Su\'arez$^{5}$,}
\newauthor{M. Suran$^{18}$, R. Szab\'o$^{20}$, K. Uytterhoeven$^{10}$, Christensen-Dalsgaard$^{21}$,}
\newauthor{H. Kjeldsen$^{21}$, D. A. Caldwell$^{22}$, F. R. Girouard$^{23}$, D. T. Sanderfer$^{24}$}
\\
\\
$^{1}$South African Astronomical Observatory, P.O. Box 9, Observatory 7935, Cape Town, South Africa\\
$^{2}$Instytut Astronomiczny, Uniwersytet Wroc{\l}awski, Kopernika 11, 51-622 Wroc{\l}aw, Poland\\
$^{3}$Koninklijke Sterrenwacht van Belgi\"e, Ringlaan 3, B-1180 Brussels, Belgium\\
$^{4}$Institut f\"{u}r Astronomie, Universit\"{a}t Wien, Vienna, Austria\\
$^{5}$Instituto de Astrof\'{\i}sica de Andaluc\'{\i}a, CSIC, P.O. Box 3004, 18080 Granada, Spain\\
$^{6}$Department of Physics, University of Johannesburg, South Africa\\
$^{7}$School of Physics, University of the Witwatersrand, Johannesburg, South Africa\\
$^{8}$Instituut voor Sterrenkunde, Katholieke Universiteit Leuven, Celestijnenlaan 200D, B-3001 Leuven, Belgium\\
$^{9}$Department of Physics and Astronomy, The College of Charleston, Charleston, SC 29424, USA\\
$^{10}$Laboratoire AIM, CEA/DSM-CNRS-Universit\'e Paris Diderot; IRFU/SAp, Centre de Saclay, 91191, Gif-sur-Yvette, France\\
$^{11}$Steward Observatory, University of Arizona, 933 N. Cherry Ave., Tucson, AZ 85721, USA\\
$^{12}$Dr. Karl Remeis-Observatory \& ECAP, Astronomical Inst., FAU Erlangen-Nuremberg, Sternwartstr. 7, 96049 Bamberg, Germany\\
$^{13}$Department of Physics and Astronomy, Iowa State University, Ames, USA\\
$^{14}$Th\"{u}ringer Landessternwarte, Tautenburg, Germany\\
$^{15}$LESIA, Observatoire de Paris, CNRS, UPMC, Université Paris-Diderot, 92195 Meudon, France\\
$^{16}$Institut d'Astrophysique et de G\'eophysique de l'Universit\'e de Li\`{e}ge, Li\`{e}ge, Belgium\\
$^{17}$Queen Mary University of London, London, United Kingdom\\
$^{18}$Astronomical Institute of the Romanian Academy, Bucharest, Romania\\
$^{19}$Space Telescope Science Institute, Baltimore, MD 21218, USA\\ 
$^{20}$Konkoly Observatory of the Hungarian Academy of Sciences, Budapest, Hungary\\
$^{21}$Department of Physics and Astronomy, Building 1520, Aarhus University, 8000 Aarhus C, Denmark\\
$^{22}$SETI Institute/NASA Ames Research Center, Moffett Field, CA 94035 USA\\
$^{23}$Orbital Sciences Corporation/NASA Ames Research Center, Moffett Field, CA 94035 USA\\
$^{24}$NASA Ames Research Center, Moffett Field, CA 94035 USA
}
\begin{document}

\date{Accepted .... Received .....}

\pagerange{\pageref{firstpage}--\pageref{lastpage}} \pubyear{2011}

\maketitle

\label{firstpage}

\begin{abstract}
The analysis of the light curves of 48 B-type stars observed by {\it Kepler} is 
presented. Among these are 15 pulsating stars, all of which show low frequencies
characteristic of SPB stars.  Seven of these stars also show a few weak,
isolated high frequencies and they could be considered as SPB/$\beta$~Cep
hybrids.  In all cases the frequency spectra are quite different from what
is seen from ground-based observations.  We suggest that this is because
most of the low frequencies are modes of high degree which are predicted
to be unstable in models of mid-B stars.  We find that there are non-pulsating
stars within the $\beta$~Cep and SPB instability strips.  Apart from the pulsating 
stars, we can identify stars with frequency groupings similar to what is
seen in Be stars but which are not Be stars.  The origin of the groupings
is not clear, but may be related to rotation.  We find periodic variations in other 
stars which we attribute to proximity effects in binary systems or possibly 
rotational modulation.  We find no evidence for pulsating stars between the cool 
edge of the SPB and the hot edge of the $\delta$~Sct instability strips.  None 
of the stars show the broad features which can be attributed to 
stochastically-excited modes as recently proposed.  Among our sample of B stars 
are two chemically peculiar stars, one of which is a HgMn star showing rotational 
modulation in the light curve. 
\end{abstract}


\section{Introduction}
The {\it Kepler} mission is designed to detect Earth-like planets around solar-type 
stars by the transit method \citep{Koch2010}. {\it Kepler} will continuously 
monitor the brightness of over 150\,000 stars for at least 3.5~yr in a 105 
square degree fixed  field of view.  In the course of the mission, extremely accurate 
photometry will be  obtained.  The mean top-of-the-noise level in the periodogram 
(which we will call the {\it grass} level) for a star of magnitude 10 is about
3~ppm for an observing run of 30\,d, dropping to about 1~ppm after one year.  For 
a star of magnitude 12 the corresponding grass noise level is about 6~ppm for one 
month and  3~ppm after one year.  These values apply to a frequency range 0--20~d$^{-1}$, 
the noise level dropping slowly towards higher frequencies. With this unprecedented 
level of precision one might expect to discover new behaviour for many stars.

Stability analyses \citep{Cox1992, Dziembowski1993, Dziembowski1993b} show that 
an opacity  bump due to a huge number of absorption metal lines can destabilize 
low-order $p$ and  $g$ modes in $\beta$~Cephei stars and high-order $g$ modes 
in slowly-pulsating B (SPB) stars.  Driving of pulsations due to the 
$\kappa$~mechanism can only occur if certain criteria are met. One of these 
criteria is that the pulsation period is of the same order or shorter than 
the thermal timescale.  Otherwise the driving region remains in thermal 
equilibrium and cannot absorb/release the heat required for driving the 
pulsations.  Another requirement is that the pressure variation is large and 
varies only slowly within the driving region.  This criterion is satisfied 
for low radial order $p$ modes and (in some models) high-order $g$ modes with 
$l <$ 6 in $\beta$~Cep stars in the iron-opacity bump region.  In $\beta$~Cep 
stars this opacity bump is located in a relatively shallow layer where the 
thermal timescale is well below 1~d.  It turns out that the timescale constraint 
is fulfilled  only for $p$ modes of low radial order and for $g$ modes with 
$l >$ 6 \citep{Dziembowski1993}.  In less massive stars, the iron-opacity 
bump is located in a deeper layer and, in addition, the luminosity is lower.  
Both factors contribute to an increase in the thermal timescale.  Consequently, 
all $p$ modes are stable because their periods are much shorter than the 
thermal timescale, while high radial order $g$ modes with low $l$ fulfill 
the timescale and pressure amplitude constraints and can now be driven.  Thus 
we find two instability regions among the B stars: the high-luminosity $\beta$~Cep 
instability strip and the low-luminosity SPB instability strip.

A number of $\beta$~Cep stars, however, show low-frequency pulsations 
characteristic of SPB stars \cite[see e.g.,][]{LeContel2001A&A...380..277L, 
Chapellier2006A&A...448..697C, deCat2007, Pigulski2008b, Pigulski2008, 
Handler2009}. These are known as  $\beta$~Cep/SPB hybrids.  In general, high 
radial order $g$ modes are heavily damped in B stars of high mass because they 
have high amplitudes in the deep layers just above the convective core.  The 
temperature variation gradient in this region is very large, leading to 
significant heat loss and damping of the pulsations.  However, the structure 
of the eigenfunctions plays an important role and for some modes driving in 
the iron-opacity bump region exceeds damping in the inner layers.  Hence 
some low-frequency $g$ modes can occur together with high frequency $p$ 
modes in the  same star.

Unfortunately, the $g$ modes observed in some $\beta$~Cep stars are usually 
found to be stable in the models \citep[e.g.][]{Dziembowski2008}.  It is also 
found that the observed range of instability at high frequencies is wider 
than can be explained.  
The problem may be resolved by postulating an overabundance of iron-peak 
elements in the driving region by up to a factor of four \citep{Pamyatnykh2004}.  
The use of OP opacities and the recently-revised solar mixture \citep{Asplund2009} 
requires a modest 50 percent enhancement of Fe in the iron-opacity bump for 
$\nu$~Eri models \citep{Zdravkov2008}.  Since there is no observational evidence 
to suggest that the current metal abundances are in error by that amount, the 
required opacity enhancement needs the increase in metal abundance to be 
confined to the iron-opacity bump region.  Unfortunately, any enhancement of 
Fe in this region cannot be hidden and should be detectable in the photosphere 
\citep{Seaton1999}.  Either a further upwards revision of the opacities is 
required or some physics is missing from model calculations.

The hybrid stars provide an important test of the models because they probe a 
very wide frequency range.  Hybrid behaviour is more common in models using 
OP opacities than OPAL opacities \citep{Miglio2007}.  This is mainly due to 
the fact that the metal bump opacity in the OP data occurs at a deeper level 
where the temperature is about 20\,000~K higher than in OPAL.  Also, the period 
gap between $\beta$~Cep stars and SPB stars is expected to decrease as $l$ 
increases \citep{Balona1999}.  If most of the low-frequency modes in $\beta$~Cep 
stars have high $l$, it might explain why such modes have low amplitudes 
which are difficult to detect from the ground but are seen in space observations 
of these stars \cite[e.g.][]{Degroote2009b}.

In the past, model calculations for $\beta$~Cep and SPB stars have been mostly 
confined to modes of low spherical harmonic degree, $l \le$ 2. With the 
precision attainable by space observations, it is expected that modes of much 
higher values of $l$ should be detected.  One may therefore expect that the 
view of $\beta$~Cep/SPB pulsations from space may differ considerably from the 
view obtained by ground-based observations. This is indeed the case, as will be 
shown.

The iron-opacity peak also leads to the formation of a small convective zone in 
the envelope of sufficiently luminous OB stars \citep{Stothers1993}.  It is often 
accompanied by an even smaller convective zone due to partial ionization of helium.
These convective zones are so thin that they have no influence on the pulsational 
stability or energy transport.  The convection zone due to the iron-opacity 
bump is more prominent for lower surface gravity, higher luminosity and higher 
initial metallicity \citep{Cantiello2009}.  These thin convective zones may be 
connected to photospheric turbulence (as seen, for example in the broad spectral 
lines in O-type stars), spectral line variability, wind clumping and the generation 
of a magnetic field.

In this context, one of the most interesting and important results is that obtained 
from {\it CoRoT} observations of the $\beta$~Cep star HD\,180642 (V1449~Aql).
In addition to a bulk of coherent modes with stable amplitudes, broad structures 
in the frequency range 9--22~d$^{-1}$ were detected \citep{Belkacem2009}.
These structures have been suggested to be modes with short lifetimes which are 
stochastically excited in a thin convective envelope.  In addition, {\it CoRoT} 
observations of the O-type star HD\,46149 reveal a pattern characteristic of 
stochastic oscillations, which are predicted to be similarly excited as in B 
stars due to thin convective layers in the envelope \citep{Degroote2010A&A...519A..38D}.
With the outstanding precision of {\it Kepler} observations, it is possible
to search for the presence of stochastically-excited modes in a larger sample 
of early B-type stars. 

Finally, of course, there is the prospect for uncovering hitherto unsuspected 
types of variability among B stars.  While it is not possible to deduce the 
nature of stars solely from an analysis of the light curves, one may obtain 
some clues by classifying stars into similar groups from which a common mechanism 
may, with further observations, be deduced.  In this paper we confirm some of the 
results obtained by {\it CoRoT} \citep{Degroote2009}, though testing of our 
speculations must await further observations.

In this paper we review the photometric variability of all {\it Kepler} 
Asteroseismic Science Consortium (KASC) targets that were identified as main 
sequence B-type stars.  The paper is organized as follows.
Section~\ref{MSB} describes the selection of B-type stars from the {\it Kepler} 
photometry database.  Section~\ref{fundpar} shows how stellar
parameters were derived for the selected sample.  Section~\ref{freqan} 
describes the time-series analysis followed by a presentation and review of 
the different types of variability among the selected sample (Section~\ref{notes}).
In Section~\ref{comparison} we discuss the SPB and $\beta$~Cep variables that have 
been identified and compare their properties with models.  The issue of 
stochastically-excited modes in early B-type stars is discussed in 
Section~\ref{stochastic}.

\section{Selection of main-sequence B-type stars}
\label{MSB}
Main sequence B-type stars (MSB stars) are members of Population I which are 
confined to the Galactic disk.  Consequently, very few such stars are expected 
to be located at distances larger than about 100~pc above and below the Galactic 
plane.  The {\it Kepler} field of view lies within Galactic latitudes 6$\degr < 
b <$ 20$\degr$.  A mid-B star with an apparent magnitude of $V =$ 10~mag at 
these latitudes will be about 600 pc above the Galactic plane at a distance of 
about 4~kpc.  Clearly, only the nearest and brightest B stars are expected to 
be included in the {\it Kepler} field of view.  Only some of the brightest 
stars in this field have been observed spectroscopically, which makes it very 
difficult to select the MSB stars.

\begin{figure}
\centering
\includegraphics[width=84mm]{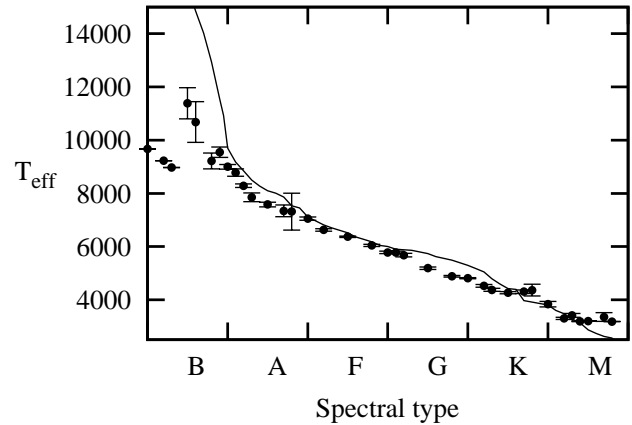}
 \caption{The effective temperatures from the KIC catalogue as a function
of spectral type. The line is a recent effective temperature --
spectral type calibration \citep{Mamajek2010}.}
\label{fig:temp}
\end{figure}

Fortunately, in preparation for the mission, Sloan Digital Sky Survey (SDSS)-like 
$griz$, D51 (510~nm) and {\sc 2MASS} $JHK$ photometry was obtained for most stars 
\citep{Batalha2010}. Effective temperatures, surface gravities and radii derived 
from this photometry are included in the {\it Kepler} Input Catalogue (KIC) 
\footnote{http://archive.scsi.edu/kepler.}.  It should be noted that the above 
filters do not include the Balmer-jump band which is essential for the 
characterization of hot objects.  Since effective temperatures are crucial for 
selecting B-type stars, we plotted the effective temperatures for stars with 
known spectral types to test the validity of the KIC calibration.  The results 
shown in Fig.~\ref{fig:temp} indicate that the KIC temperatures are not very 
reliable for B stars and are usually largely underestimated \citep[see also][]
{Molenda2010, Lehmann2010}.  The recently released {\it GALEX} database 
({\tt http://galex.stsci.edu/GR6/}) can also be useful in distinguishing between 
B-type and later stars, although there is, as yet, few matches with B stars in 
the {\it Kepler} field (see also Section\,\ref{par:sed}).

In order to maximize our chances of finding main sequence B-type stars in the 
{\it Kepler} database, we used the following selection criteria: (i) all stars 
classified as B type in the HD or other catalogues and (ii) all stars with 
effective temperatures greater than 10\,000~K in the KIC.  This was supplemented 
by recent spectroscopic investigations of stars in the {\it Kepler} field, such 
as those of \cite{Lehmann2010} and \cite{Catanzaro2010}.  These selection criteria 
will inevitably lead to the inclusion of hot compact stars, mainly hot subdwarfs 
and white dwarfs.  We therefore used the list of compact objects compiled by 
\cite{Ostensen2010} and eliminated all objects that were classified by these 
authors as non-main sequence stars.  Several objects were excluded because their 
effective temperature derived by \cite{Lehmann2010} appeared to be smaller than 
10\,000~K.  These criteria resulted in a list of 72 presumably MSB stars.
\begin{figure}
\centering
\includegraphics[width=84mm]{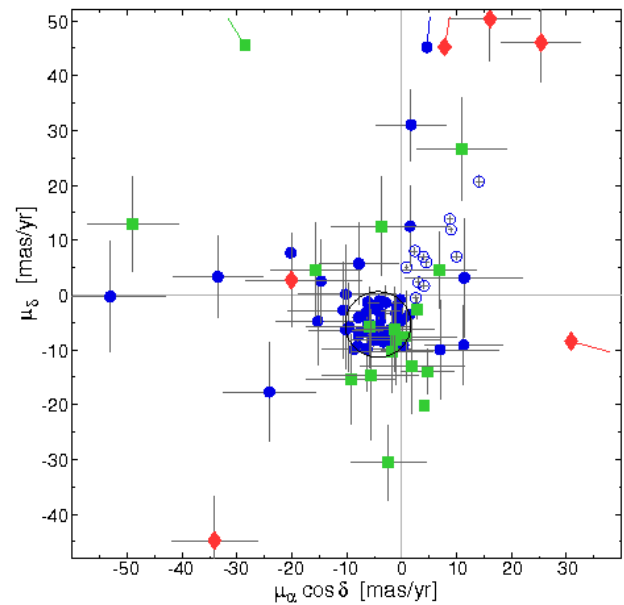}
 \caption{The UCAC3 proper motions of a selected sample of presumably MSB stars
(filled and open circles).  For comparison, the location of excluded hot subdwarfs 
(squares)  and white dwarfs (diamonds) is shown.  The eleven MSB stars which are 
likely members  of Gould Belt are shown with open circles.  Four stars that lie 
outside the  figure are shown with appropriate symbols. The large circle with a 
radius of  6~mas\,yr$^{-1}$ is centered at $((\mu_\alpha\cos\delta)_0$, 
$\mu_{\delta,0}$) $=$  ($-$4.3,$-$5.4) mas\,yr$^{-1}$.}
\label{fig:pm}
\end{figure}

Proper motions can be used to further refine our sample of stars.
We used proper motions from the UCAC3 catalogue \citep{Zacharias2009} to 
identify and exclude probable non-MSB stars.  All but seven stars in our 
sample are listed in this catalogue.  Fig.~\ref{fig:pm} shows the location of 
the selected stars in the proper-motion diagram.  As can be seen from the 
figure the MSB stars, especially the brightest stars (with small error bars), 
form a well-defined clump centered at $((\mu_\alpha \cos\delta)_0$, 
$\mu_{\delta,0}$) $=$ ($-$4.3,$-$5.4) mas\,yr$^{-1}$.  The center of the clump 
is consistent with the average proper motion of the young population in the 
direction of the {\it Kepler} field of view as can be judged from proper 
motions of open clusters \citep{Kharchenko2003}.  However, there is a sample of 
eleven bright stars, shown with open circles in Fig.~\ref{fig:pm}, which fall 
well outside the clump but are classified as main sequence B8 or B9 stars.
Their magnitudes and proper motions are consistent with membership of the 
Gould Belt which happens to cross the {\it Kepler} field of view 
\citep[see][and references therein]{Elias2009}.

\begin{figure}
\centering
\includegraphics[width=84mm]{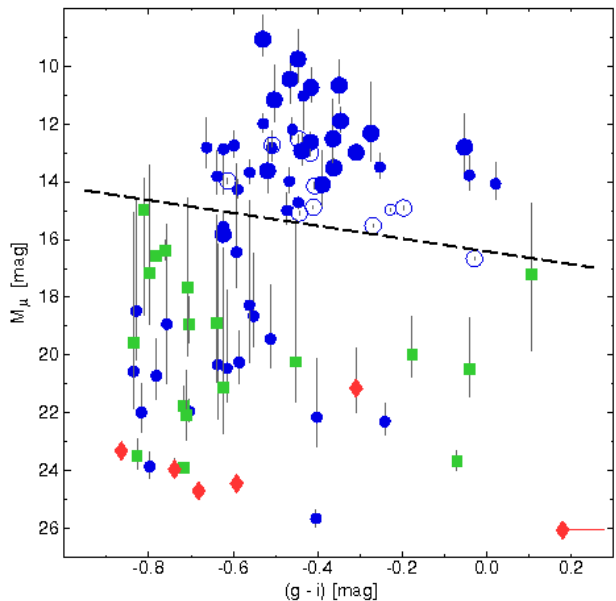}
 \caption{The $M_\mu$ parameter as a function of $(g-i)$ colour index for a
 sample of MSB stars, white dwarfs and subdwarfs from the list of {\O}stensen 
 et al.~(2010).  The symbols have the same meaning as in Fig.~\ref{fig:pm}.
 Large symbols denote MSB stars with known MK spectral types.  The dashed line,
 with the adopted ratio $A_g/E(g-i) = 2.222$ is drawn to separate MSB stars from 
 compact stars.}
\label{fig:rpm}
\end{figure}

The three groups of hot stars we are considering here (MSB stars, hot subdwarfs 
and white dwarfs) show different spreads in proper motions (Fig.~\ref{fig:pm}).
In particular, white dwarfs have the largest proper motions.  The proper motions 
of hot subdwarfs are smaller, but still larger than for MSB stars.
However, proper motions alone are not sufficient to distinguish between the 
three groups.  For example, some hot subdwarfs are to be found among the clump 
of MSB stars in Fig.~\ref{fig:pm}.  Statistically, the larger the proper motion, 
the nearer the star.  We may therefore use proper motion, $\mu$, as an 
approximate indicator of the distance, $D$.  From this, we may define a 
quantity which is related to the absolute magnitude.  This is useful because 
it is the absolute magnitude which best discriminates between the three groups 
of stars.

We may define a parameter $M_\mu$ as an indicator of the absolute magnitude of 
a star: $$M_\mu = g + \mbox{5}\log(\Delta\mu/(\mbox{mas yr$^{-1}$})),$$ where 
$g$ is a Sloan-like $g$ magnitude taken from the KIC catalogue and $\Delta\mu$ 
is the difference in proper motion from the center of the clump in Fig.~\ref{fig:pm}. 
In order to avoid singularity in calculating $M_\mu$, the lower limit for 
$\Delta\mu$ was set to 1~mas\,yr$^{-1}$.  The definition of $M_\mu$ is a slight 
modification of the parameter called ``reduced proper motion'' by \cite{Ostensen2010} 
for a similar purpose.  A plot of $M_\mu$ as a function of the $(g-i)$ index 
from the KIC is shown in Fig.~\ref{fig:rpm}.  It can be seen that the three 
groups are quite well separated by $M_\mu$.  In particular, the dashed line 
in Fig.~\ref{fig:rpm} separates MSB stars (above the line) from hot subdwarfs 
(below the line).  The separation is verified by the location of MSB stars with 
good MK spectral classification which we know are certainly not compact objects 
(larger symbols in Fig.~\ref{fig:rpm}).  Stars belonging to Gould Belt have 
different proper motions and are located far from the clump of MSB stars in 
Fig.~\ref{fig:pm}.  It is therefore not surprising that they might have a larger 
$M_\mu$ and indeed, KIC\,11973705 which is a known MSB star (see
Table\,\ref{tab:tar}), lies on the fringe of the Gould Belt distribution and 
below the dashed line.  The spectrum of KIC\,12258330, another star below the 
line, shows that it is an MSB star (Table\,\ref{tab:tar}).

All stars falling below the dashed line in Fig.~\ref{fig:rpm} except for these 
two stars were excluded from our sample as they are unlikely to be MSB stars.
In addition, we excluded seven very faint stars with $g$ magnitudes in the 
range 14.5--17.0~mag which do not have proper motions in the UCAC3 catalogue.
It is unlikely they are MSB stars as they would lie at a very great distance 
from the Galactic plane.  We are then left with a sample of 48 MSB stars for 
detailed frequency analysis.  These stars are listed in Table \ref{tab:tar}.
We have to stress, however, that even though the selection of our target stars 
was done in a consistent way, we cannot exclude the possibility that some MSB 
stars might have been overlooked.
  
\begin{table*}
\caption{List of 48 main sequence B-type stars discussed in this paper.
The duration of the short-cadence (SC) and long-cadence (LC) runs for each 
star is given.  The SDSS $g-i$ index and the $M_\mu$ parameter is discussed in 
the text.  The notation FG means a star with frequency groupings while Rot
implies the variability could be due to rotation.
}
\label{tab:tar}
\begin{tabular*}{1.0\textwidth}
{@{\extracolsep{\fill}}r 
@{\extracolsep{\fill}}r 
@{\extracolsep{\fill}}r 
@{\extracolsep{2mm}}r 
@{\extracolsep{\fill}}r 
@{\extracolsep{\fill}}c
@{\extracolsep{\fill}}c 
@{\extracolsep{\fill}}c 
@{\extracolsep{1mm}}l 
@{\extracolsep{\fill}}l} 
\hline\noalign{\smallskip}
\multicolumn{1}{c}{KIC} & 
\multicolumn{1}{c}{2MASS} & 
\multicolumn{1}{c}{SC} & 
\multicolumn{1}{c}{LC} & 
\multicolumn{1}{c}{$g$~$^a$} & 
\multicolumn{1}{c}{$(g-i)$~$^a$} & 
\multicolumn{1}{c}{$M_\mu$} & 
\multicolumn{1}{c}{Classification} &
\multicolumn{1}{l}{Sp.~type} & 
\multicolumn{1}{l}{Notes} 
\\
\multicolumn{1}{c}{number} &
\multicolumn{1}{c}{number} &
\multicolumn{1}{c}{[d]} &
\multicolumn{1}{c}{[d]} & 
\multicolumn{1}{c}{[mag]} & 
\multicolumn{1}{c}{[mag]} & 
\multicolumn{1}{c}{[mag]} & 
\multicolumn{1}{c}{} &
\multicolumn{1}{c}{} &
\multicolumn{1}{c}{}
\\
\noalign{\smallskip}
\hline\noalign{\smallskip}
 3240411 & ~19310672+3819417 &   9.7 & 322.1 & 10.08 &   ---   & ---   &Hybrid& B2\,V\,$^b$                     & \\                                          
 3756031 & ~19350581+3852500 &  ---  & 321.6 & 9.86  & $-$0.47 & 10.44 &SPB + $\delta$\,Sct?& B5\,V-IV\,$^b$    & \\                            
 3839930 & ~19130682+3859204 &  ---  & 321.6 & 10.68 & $-$0.51 & 12.82 &SPB& ---                                & \\                                             
 3848385 & ~19233702+3859363 &  ---  &  44.4 &  8.86 & $-$0.35 & 10.65 & Rot + ?& B8\,V\,$^c$                   &  GALEX mags. \\                                        
 3865742 & ~19411999+3856436 &  ---  & 321.6 & 10.96 & $-$0.43 & 11.03 &FG& B                                   & \\                                            
 4276892 & ~19400775+3920382 &  25.9 &  44.4 & 9.06  & $-$0.35 & 11.89 &Bin + ?& B8--A0\,$^d$                   & SB1?\,$^d$, FUV GALEX mag.\\               
 4581434 & ~19455619+3939557 &  30.0 &  33.4 & 9.12  & $-$0.20 & 14.90 &Bin/Rot& B9--A5\,$^d$                   & SB1\,$^d$\\                                  
 5130305 & ~19563429+4014532 &  ---  & 321.6 & 10.07 & $-$0.36 & 13.51 &Rot& B9\,IV-V\,$^b$                     & \\                                             
 5217845 & ~19555093+4023268 &  ---  & 321.6 & 9.35  & $-$0.27 & 12.32 &Bin + Rot& B8.5\,III\,$^b$              & binary\,$^b$ \\                            
 5304891 & ~19551942+4029573 &  30.0 &  44.4 & 9.07  & $-$0.31 & 12.98 &Rot& B4--B8\,$^d$                       & \\                                             
 5458880 & ~19390756+4037365 &  28.9 & 311.1 & 7.63  & $-$0.53 & ~~9.05 &Bin/Rot& B0\,III\,$^e$, B0.5\,II\,$^f$ & \\ 
 5479821 & ~19572477+4038084 &  ---  & 321.6 & 9.74  & $-$0.42 & 12.63 &Bin + Rot/?& B5.5\,V\,$^b$              & \\                                     
 5786771 & ~19214208+4105482 &  30.0 &  44.4 & 9.01  & $-$0.41 & 14.13 &Rot& B9--A5\,$^d$                       & \\                                             
 6128830 & ~19334968+4128452 &  28.9 &  44.4 & 9.09  & $-$0.46 & 12.19 &Rot& B6\,HgMn\,$^d$                     & \\                                             
 6954726 & ~19344680+4226081 &  ---  & 321.6 &11.72  & $-$0.52 & 13.62 &FG& Be\,$^{g,h}$                        & Be star, StH$\alpha$166 \\                    
 7599132 & ~19155121+4312084 &  33.4 & 321.6 &  9.22 & $-$0.44 & 15.08 &Bin& B8.5\,V\,$^b$                      & \\                                             
 7974841 &~ 19463656+4345480 &  27.1 &  44.4 & 8.05  & $-$0.41 & 14.88 &Rot/SPB& B8\,V\,$^i$                    & \\ 
 8018827 & ~19115652+4353258 &  33.4 & ---   & 7.90  & $-$0.45 & 12.52 &Rot& B9\,$^j$                           & FUV GALEX mag. \\                                             
 8057661 & ~20001799+4349558 &  ---  & 321.6 & 11.62 & $-$0.05 & 12.80 &Hybrid& OB$-$H\,$^k$                    & Be star, KW97 41-38 \\                      
 8087269 & ~19110975+4359571 &  30.0 & 321.6 & 11.94 & $-$0.47 & 14.99 &FG&                                     & GALEX mags.\\                                  
 8161798 & ~19213515+4403022 &  ---  & 321.6 & 10.33 & $-$0.44 & 12.94 &Bin& B8\,$^j$                           & GALEX mags. \\                                             
 8177087 & ~19422573+4401183 &  33.4 & 311.1 & 7.90  & $-$0.50 & 11.16 &SPB& B5\,V\,$^i$, B7\,III\,\,$^b$       & \\ 
 8324268 & ~19565014+4416159 &  ---  & 311.1 & 7.81  & $-$0.42 & 13.00 &Rot& B9\,V\,$^f$, B8p\,Si\,$^l$         & V2095\,Cyg, $\alpha^2$\,CVn star \\            
 8351193 & ~19012177+4423115 &  30.3 &  33.4 & 7.40  & $-$0.51 & 12.73 &Rot& B9\,V\,$^i$                        & \\                                             
 8381949 & ~19484697+4421208 &  ---  & 321.6 & 10.97 & $-$0.36 & 12.51 &Hybrid& OB$-^k$                         & \\ 
 8389948 & ~19565533+4420307 &  31.1 & 321.6 & 9.13  & $-$0.27 & 15.52 &Rot& B9.5\,V-IV\,$^b$                   & \\ 
 8459899 & ~20021338+4426469 &  30.7 &  44.4 & 8.58  & $-$0.42 & 10.73 &SPB& B2\,III\,$^f$, B4.5\,IV\,$^b$      & SB2?\,$^b$ \\ 
 8488717 & ~19172024+4433508 &  ---  & 321.6 & 11.62 & $-$0.39 & 14.09 &Rot/Bin& B9\,$^j$                       & \\                                         
 8692626 & ~19324040+4452514 &  30.0 &  44.4 & 8.27  & $-$0.23 & 14.96 &Rot + ?& [B9]\,$^m$                     & \\                                         
 8714886 & ~19591018+4451448 &  ---  & 321.6 & 10.86 & $-$0.25 & 13.49 &Hybrid/FG &                             & GALEX mags. \\                                     
 8766405 & ~19455212+4455071 &  33.4 & 321.6 & 8.71  & $-$0.45 & ~~9.74  &FG& B5\,V\,$^f$, B7\,III\,$^b$        & SB2?\,$^b$ \\                                   
 9655433 & ~19372486+4618390 &  ---  & 311.1 & 11.93$^*$ & --- & ---   &$\delta$~Sct& A3:\,$^n$, B7\,$^o$       & NGC\,6811 \#114, B66 \\               
 9716301 & ~19371376+4625257 &  27.1 & 321.6 & 11.68 & $+$0.02 & 14.07 &FG& A3\,$^n$, B9\,$^o$                  & NGC\,6811 \#34, B22  \\                       
 9716456 & ~19372852+4624183 &  ---  & 321.6 & 12.00 & $-$0.04 & 13.77 &Bin + Rot& A1\,$^n$, B9\,$^o$           & NGC\,6811 \#99, B29 \\                   
 9964614 & ~19480282+4653445 &  ---  & 321.6 & 10.70 & $-$0.45 & 14.72 &Hybrid& ---                             & \\                                          
10130954 & ~19102372+4709441 &   9.7 & 322.1 & 10.85 & $-$0.66 & 12.81 &Bin/Rot& B\,$^p$                        & FUV GALEX mag. \\ 
10285114 & ~19440272+4721169 &   9.7 & 322.1 & 10.99 & $-$0.59 & 14.27 &FG& B\,$^p$                             & \\                                            
10536147 & ~19291192+4744430 & 322.1 & 322.1 & 11.33 & $-$0.64 & 13.81 &Hybrid& ---                             & \\                                          
10658302 & ~19150768+4754190 &  30.7 &  44.4 & 12.86 & $-$0.62 & 12.86 &SPB/FG& B\,$^p$                         & GALEX mags. \\ 
10797526 & ~19274936+4810363 & 110.2 &  ---  & 8.09  & $-$0.61 & 13.96 &SPB& OB$-\,^q$                          & \\ 
10960750 & ~18522875+4824141 &  27.1 & 321.6 & 9.67  & ---     & ---   &Hybrid& B4\,V:\,$^r$, B2.5\,V\,$^b$     & SB1?\,$^d$\\ 
11360704 & ~19460296+4907596 &  ---  & 321.6 & 10.51 & $-$0.56 & 13.68 &FG& ---                                 & \\                                            
11454304 & ~19252435+4918562 &   9.7 &  33.4 & 12.69 & $-$0.62 & 15.55 &SPB& B\,$^p$                            & GALEX mags. \\                                             
11817929 & ~19353836+5002354 &   9.7 &  33.4 & 10.15 & $-$0.60 & 12.74 &cst& B\,$^p$                            & \\                                             
11973705 & ~19464258+5021013 &   9.7 & 322.1 & 9.15  & $-$0.03 & 16.67 &Bin + SPB + $\delta$\,Sct& B8.5\,V-IV\,$^b$  & SB2\,$^b$, $\delta$~Sct \\ 
12207099 & ~19221039+5048401 &  ---  & 321.6 & 10.11 & $-$0.47 & 13.99 &Bin& B9\,III-II\,$^b$                   & SB2?\,$^b$, GALEX mags. \\                       
12217324 & ~19440458+5053578 &  30.0 &  44.4 & 8.10  & $-$0.53 & 11.98 &Rot& A0\,$^m$                           & \\                                             
12258330 & ~19255755+5054033 &  30.7 & 321.6 & 9.25  & $-$0.62 & 15.82 &FG& B5\,V\,$^r$, B5.5\,V-IV\,$^b$       & \\                                            
\hline
\end{tabular*}

{\small Source of information: 
$^a$\,KIC, 
$^b$\,\cite{Lehmann2010}, 
$^c$\,\cite{Sato1990}, 
$^d$\,\cite{Catanzaro2010},
$^e$\,\cite{Guetter1968},
$^f$\,\cite{Hill1977},
$^g$\,\cite{Stephenson1986},
$^h$\,\cite{Downes1988},
$^i$\,\cite{Bartaya1983},
$^j$\,\cite{Macrae1952}
$^k$\,\cite{Hardorp1964},
$^l$\,\cite{Grenier1999},
$^m$\,HD catalogue,
$^n$\,\cite{Becker1947},
$^o$\,\cite{Lindoff1972},
$^p$\,\cite{Ostensen2010},
$^q$\,\cite{Svolopoulos1969},
$^r$\,\cite{Dworetsky1982},\\
}
{\small Note: $^*$ {\it Kepler} magnitude.}
\end{table*}

\section{Determination of fundamental parameters}
\label{fundpar}

\begin{table*}
\caption{The determination of the stellar parameters (effective temperature 
T$_{\rm eff}$, surface gravity $\log g$, projected rotational velocity 
$v \sin i$ and reddening $E(B-V)$) for the objects of Table\,\ref{tab:tar} 
for which ground-based follow-up observations are available.  Errors in these 
parameters are given in brackets.  We give the results obtained from spectroscopy 
in columns 3--5, from Str\"omgren photometry in columns 6--7 and from spectral 
energy distribution fitting in columns 8--9.  For each object, the 
$\log (\rm{L/L}_{\odot})$ given in column 9 was derived from the leftmost values of 
T$_{\rm eff}$ and $\log g$.
}
\label{tab:param}
\begin{tabular*}{0.875\textwidth}{c@{\extracolsep{\fill}}ccc@{\extracolsep{\fill}}cc@{\extracolsep{\fill}}ll@{\extracolsep{\fill}}c}
\hline\noalign{\smallskip}
         &\multicolumn{3}{c}{Spectroscopy}&\multicolumn{2}{c}{Str\"omgren photometry}&\multicolumn{2}{c}{SED fitting}& \\ 
         \noalign{\smallskip}\cline{2-4}\cline{5-6}\cline{7-8}\noalign{\smallskip}
KIC  & T$_{\rm eff}$       & $\log g$ (cgs)   & $v \sin i$&T$_{\rm eff}$&$\log g$ (cgs)&
\multicolumn{1}{c}{T$_{\rm eff}$}&\multicolumn{1}{c}{$E(B-V)$}&$\log(\rm{L/L}_{\odot})$\\ 
number    & [K]                 & [dex]         & [km\,s$^{-1}$] &[K]          &[dex]   &\multicolumn{1}{c}{[K]} &  \multicolumn{1}{c}{[mag]} &[dex]   \\ 
\noalign{\smallskip}\hline\noalign{\smallskip}                                 				               	       	
\noalign{\smallskip}\hline\noalign{\smallskip}                                 				               	       	          
  ~~3240411 & 20\,980(880)\,$^a$ & 4.01(12)\,$^a$  &   43(5)\,$^a$  & 20\,550  &  3.94  & 21\,000(8\,800) & 0.09(10)  &  3.56  \\
  ~~3756031 & 15\,980(310)\,$^a$ & 3.75(6)\,$^a$   &   31(4)\,$^a$  & 16\,310  &  4.19  & 16\,000(1\,600) & 0.10(3)   &  3.21  \\
  ~~3839930 & 16\,500\,$^c$      & 4.2\,$^c$       &  ---           & 17\,160  &  4.51  & 18\,200(7\,000) & 0.10(9)   &  2.75  \\
  ~~3848385 &    ---             & ---             &  ---           & 11\,680  &  3.76  & 12\,200(800)    & 0.13(3)   &  2.37  \\
  ~~3865742 & 19\,500\,$^c$      & 3.7\,$^c$       &  133(10)\,$^d$ & 20\,190  &  4.20  & 19\,400(1\,800) & 0.17(2)   &  3.76  \\
  ~~4276892 & 10\,800(600)\,$^b$ & 4.1(2)\,$^b$    &   10\,$^b$     &~~9\,510  &  4.21  & 10\,400(600)    & 0.11(4)   &  1.87  \\
  ~~4581434 & 10\,200(200)\,$^b$ & 4.2(2)\,$^b$    &  200\,$^b$     &~~9\,130  &  4.43  & 10\,200(2\,200) & 0.17(17)  &  1.62  \\
  ~~5130305 & 10\,670(200)\,$^a$ & 3.86(7)\,$^a$   &  155(13)\,$^a$ & 10\,190  &  4.38  & 11\,000(600)    & 0.11(3)   &  2.13  \\
  ~~5217845 & 11\,790(260)\,$^a$ & 3.41(10)\,$^a$  &  237(16)\,$^a$ & 11\,740  &  3.76  & 12\,000(600)    & 0.19(2)   &  2.92  \\
  ~~5304891 & 13\,100(700)\,$^b$ & 3.9(2)\,$^b$    &  180\,$^b$     & 11\,530  &  3.81  & 11\,800(600)    & 0.14(2)   &  2.56  \\
  ~~5458880 &    ---             & ---             &  ---           & 24\,070  &  3.18  & 27\,600(5\,400) & 0.18(5)   &  4.65  \\
  ~~5479821 & 14\,810(350)\,$^a$ & 3.97(9)\,$^a$   &   85(8)\,$^a$  & 17\,580  &  4.43  & 17\,600(3\,400) & 0.16(5)   &  2.76  \\
  ~~5786771 & 10\,700(500)\,$^b$ & 4.2(2)\,$^b$    &  200\,$^b$     &~~9\,630  &  4.39  & 10\,600(1\,000) & 0.06(6)   &  1.73  \\
  ~~6128830 & 12\,600(600)\,$^b$ & 3.5(3)\,$^b$    &   15\,$^b$     &  ---     &        &        ---      & ---       &  2.96  \\
  ~~6954726 &    ---             & ---             &  160\,$^d$     &  ---     &        &        ---      & ---       &  ---   \\
  ~~7599132 & 11\,090(140)\,$^a$ & 4.08(6)\,$^a$   &   63(5)\,$^a$  & 10\,560  &  4.39  & 11\,600(2\,200) & 0.13(11)  &  1.95  \\
  ~~7974841 &    ---             & ---             &  ---           & 10\,820  &  4.41  & 11\,800(1\,200) & 0.13(5)   &  1.85  \\
  ~~8018827 &    ---             & ---             &  ---           & 10\,410  &  4.37  & 11\,200(1\,000) & 0.10(5)   &  1.79  \\
  ~~8057661 &    ---             & ---             &   49(5)\,$^d$  & 21\,360  &  4.23  & 20\,600(2\,800) & 0.40(3)   &  3.55  \\
  ~~8087269 & 14\,500\,$^c$      & 3.9\,$^c$       &  271(10)\,$^d$ &  ---     &        &        ---      & ---       &  2.80  \\
  ~~8161798 & 12\,300\,$^c$      & 4.0\,$^c$       &  ---           & 12\,460  &  4.61  & 12\,200(800)    & 0.08(3)   &  2.29  \\
  ~~8177087 & 13\,330(220)\,$^a$ & 3.42(6)\,$^a$   &   22(2)\,$^a$  & 13\,380  &  3.79  & 13\,400(1\,000) & 0.10(3)   &  3.20  \\
  ~~8324268 &    ---             & ---             &  ---           & 14\,330  &  4.46  & 14\,400(2\,000) & 0.12(4)   &  2.47  \\
  ~~8351193 &    ---             & ---             &  ---           & 10\,210  &  4.61  & 11\,000(400)    & 0.02(2)   &  1.64  \\
  ~~8381949 & 24\,500\,$^c$      & 4.3\,$^c$       &  ---           & 21\,000  &  3.82  & 22\,400(7\,400) & 0.24(7)   &  3.61  \\
  ~~8389948 & 10\,240(340)\,$^a$ & 3.86(12)\,$^a$  &  142(12)\,$^a$ &~~9\,690  &  4.33  & 10\,400(1\,800) & 0.16(13)  &  2.03  \\
  ~~8459899 & 15\,760(240)\,$^a$ & 3.81(5)\,$^a$   &   53(4)\,$^a$  & 15\,950  &  4.09  & 16\,000(2\,000) & 0.14(4)   &  3.10  \\
  ~~8488717 & 11\,000\,$^c$      & 4.0\,$^c$       &  ---           & 10\,160  &  4.44  & 11\,200(1\,400) & 0.10(7)   &  2.03  \\
  ~~8692626 &    ---             & ---             &  ---           &~~9\,826  &  4.71  & ~~9\,400(1\,200)& 0.12(12)  &  1.52  \\
  ~~8714886 & 19\,000\,$^c$      & 4.3\,$^c$       &  ---           & 18\,505  &  4.49  & 18\,000(3\,400) & 0.26(5)   &  2.98  \\
  ~~8766405 & 12\,930(220)\,$^a$ & 3.16(8)\,$^a$   &  240(12)\,$^a$ & 14\,050  &  3.74  & 14\,400(1\,800) & 0.11(4)   &  3.46  \\
  ~~9964614 & 20\,300\,$^c$      & 3.9\,$^c$       &  ---           & 19\,471  &  3.75  & 23\,400(7\,400) & 0.14(10)  &  3.61  \\
 10130954 & 19\,400\,$^c$        & 4.0\,$^c$       &  ---           & 18\,663  &  3.96  & 19\,600(5\,200) & 0.05(6)   &  3.38  \\
 10285114 & 18\,200\,$^c$        & 4.4\,$^c$       &  ---           & 16\,229  &  4.16  & 16\,400(4\,200) & 0.07(7)   &  2.76  \\
 10536147 & 20\,800\,$^c$        & 3.8\,$^c$       &  195(10)\,$^d$ &  ---     &        &        ---      & ---       &  3.79  \\
 10658302 & 15\,900\,$^c$        & 3.9\,$^c$       &  ---           &  ---     &        &        ---      & ---       &  3.02  \\
 10797526 &    ---               & ---             &  ---           & 20\,873  &  3.22  & 23\,600(5\,600) & 0.13(5)   &  4.15  \\
 10960750 & 19\,960(880)\,$^a$   & 3.91(11)\,$^a$  &  253(15)\,$^a$ & 20\,141  &  3.85  & 21\,800(8\,000) & 0.06(8)   &  3.56  \\
 11360704 & 20\,700\,$^c$        & 4.1\,$^c$       &  ---           & 17\,644  &  3.89  & 18\,200(5\,400) & 0.11(7)   &  3.66  \\
 11454304 & 17\,500\,$^c$        & 3.9\,$^c$       &  ---           &  ---     &        &        ---      & ---       &  3.25  \\
 11817929 & 16\,000\,$^c$        & 3.7\,$^c$       &  ---           & 12\,732  &  4.02  & 13\,000(1\,800) & 0.02(5)   &  3.28  \\
 11973705 & (11\,150)\,$^a$      &  (3.96)\,$^a$   &  103(10)\,$^a$ & 11\,898  &  4.25  & 11\,800(3\,800) & 0.31(26)  & (2.11) \\
 12207099 &$<$11\,000\,$^a$      & $<$3.1\,$^a$    &   43(5)\,$^a$  &  ---     &        &        ---      & ---       & (3.17) \\
 12217324 &    ---               & ---             &  ---           & 10\,192  &  4.23  & 11\,000(600)    & 0.03(3)   &  1.81  \\
 12258330 & 14\,700(200)\,$^a$   & 3.85(4)\,$^a$   &  130(8)\,$^a$  & 16\,436  &  4.35  & 16\,600(3\,400) & 0.05(6)   &  2.89  \\
\noalign{\smallskip}\hline
\end{tabular*}
\\
{\small Source of information: 
$^a$\,\cite{Lehmann2010}, 
$^b$\,\cite{Catanzaro2010},
$^c$\,Kitt Peak observations,
$^d$\,CFHT observation.
}
\end{table*}

Stellar parameters are required in order to place the stars of Table\,\ref{tab:tar} 
in the theoretical HR diagram.  We cannot use the KIC parameters for reasons 
already discussed, but fortunately more reliable values of the effective
temperature, T$_{\rm eff}$, and surface gravity, $\log g$, may be derived
from spectroscopy (Section\,\ref{par:spec}).  Narrow-band photometry  can also 
be used to derive the effective temperatures and luminosities (Section\,\ref{par:phot}). 
In addition, spectroscopic observations allow estimates of the projected 
rotational velocity, $v \sin i$, which can be used as a constraint in the 
interpretation of {\it Kepler} light curves.

\subsection{Spectroscopy} \label{par:spec}
In columns 2, 3 and 4 of Table\,\ref{tab:param}, we give T$_{\rm eff}$, $\log g$ 
and $v \sin i$, respectively, as derived from spectroscopic follow-up observations. 
The entries include the published results of \cite{Lehmann2010} and \cite{Catanzaro2010}.
New spectral classifications for some objects were also obtained with the B\&C 
spectrograph using a 400/mm grating ($R \simeq$ 550, $\lambda=$ 362--690~nm) 
attached to the Bok telescope at Kitt Peak observatory.   These observations 
were done on June 14--16 and July 28--30, 2008.  The T$_{\rm eff}$ and $\log g$ 
were derived by fitting model atmosphere grids to the hydrogen and helium 
lines visible in the spectra as described by \cite{Ostensen2010}.  Metal-line 
blanketed LTE models of solar composition as described in \cite{Heber2000A&A...363..198H}
were used.  Formal fitting errors on T$_{\rm eff}$ and $\log g$ are about 200\,K 
and  0.05\,dex respectively, while the systematic errors are probably about 5 
percent for both parameters.  Because of the low spectral resolution, the 
projected rotational velocity could not be determined and $v \sin i =$ 0 
was assumed during the fitting procedure.

For five targets (KIC\,3865742, 6954726, 8057661, 8087269 and 10536147), at 
least one high-resolution spectrum ($R =$ 81\,000, $\lambda =$ 370--1000~nm) 
was obtained with the Echelle SpectroPolarimetric Device for the Observation of 
Stars (ESPaDOnS) attached to the Canada-France-Hawaii Telescope (CFHT) at 
Mauna Kea (Hawaii) on 2010 July 25, 31 and 2010 August 4.  The {\it Kepler} 
light curves of the majority of these objects show characteristics similar to 
those of Be stars (cf. Section\,\ref{Be-like}).  Apart from the confirmed Be 
star KIC\,6954726, none of these stars showed emission lines.  These spectra 
were used only to derive a $v \sin i$ value by applying the Fourier transform 
method  to the He\,I, Mg\,II and Si\,III lines (\citealt{Gray1982ApJ...255..200G}, 
see also \citealt{SimonDiaz2007A&A...468.1063S} for a recent application to 
OB-type stars).  A full analysis of these spectra, together with a detailed 
analysis of their {\it Kepler} light curves, will be presented elsewhere 
(Guti\'errez-Soto, in preparation).

\subsection{Photometry} \label{par:phot}
\subsubsection{Str\"omgren photometry} \label{par:strom}
Standard Str\"omgren-Crawford $uvby\beta$ photometry was obtained with the 
2.1-m telescope at McDonald Observatory.  A standard two-channel photoelectric 
photometer was used in 2010 August/September in single-channel mode.
Apertures of 14.5 and 29$^{\prime\prime}$ were used depending on the brightness of 
the target and sky background.  As the filter wheel can only hold four filters 
at once, $uvby$ measurements were obtained at different times from $\beta$
observations.

A set of standard stars, selected to span the whole parameter range of the 
targets in terms of $(b-y)$, $m_1$, $c_1$, $\beta$ and $E(b-y)$,  were observed 
for transforming the measurements into the standard system.  Transformation 
equations, used in the context of a larger observing program, were adopted 
from Handler (2010, in preparation).   To maximize the number of objects 
for which stellar parameters can be estimated, we applied the calibration of 
\citet{Balona1994} to the de-reddened Str\"omgren data to obtain T$_{\rm eff}$ 
and $\log g$ (see columns 5 and 6 in Table\,\ref{tab:param}).  As can be 
seen from Table\,\ref{tab:param}, the results are, for the most part, in 
good agreement with the spectroscopic values.

\subsubsection{Spectra energy distribution (SED) fitting} \label{par:sed}

Values for T$_{\rm eff}$ were determined by fitting \citet{kurucz1993} 
model atmospheres with solar metallicity to the combined set of SDSS, 2MASS 
and Str\"omgren colours.  The calibration zero points for Str\"omgren and SDSS 
were derived from the dedicated Vega model from \citet{castelli1994} and the 
2MASS zero points were taken from \citet{cohen2003}.  Wherever possible, the 
photometry was extended with values from \citet{droege2006}, \citet{ofek2008} and 
\citet{thompson1978}.  A grid search was performed, within the effective temperature 
range 8\,000--30\,000~K in steps of 200~K and the surface gravity range between 
3.5 and 4.5~dex (cgs), in steps of 0.25~dex.  For each combination of T$_{\rm eff}$ 
and $\log g$, the model atmosphere was reddened according to the reddening law 
of \citet{cardelli1989} using $R_V =$ 3.1. The colour excess, $E(B-V)$, was 
determined by an iterative procedure using  $\chi^2$ minimization.  The confidence 
interval for T$_{\rm eff}$ was chosen to encompass all values for which 
$\chi^2$ was below twice the minimal value.  Equal weights were given to all
data points in the $\chi^2$ minimization procedure.  The value for $\log g$
cannot be obtained by this method.

We tried to make additional use of the near- and far-ultraviolet photometry 
from the {\it Galaxy Evolution Explorer} satellite \citep{Martin2005ApJ...619L...1M}.
The GALEX satellite was designed to image light from faint moderate- and
high-galactic latitude objects in both near- and far-UV (NUV and FUV, respectively) 
imaging cameras by means of a beam-splitter.  The latest General Release (GR6) 
delivery to the {\it Multi-Mission Archives at Space Telescope Science
Institute} (MAST)\footnote{Support for the MAST is provided by NASA through a 
grant from the Space Telescope Science Institute, which is operated by the 
Association of Universities for Research in Astronomy, Inc., under NASA 
Contract NAS5-26555.} in 2010 showed an overlap of some 60 percent of the 
{\it Kepler} field of view (FOV), which is far more than was delivered in 
previous releases.  The GALEX GR6 catalog were cross-matched with KIC stars 
at MAST for coordinate separations of less than 2.5 arcsec on the sky, resulting 
in FUV and/or NUV magnitudes for ten targets in Table \ref{tab:tar}.
The SED fitting was redone including these GALEX magnitudes but did not lead 
to better results.  Although GALEX-determined colours hold out great promise 
for differentiating among hot stars in the {\it Kepler} field, so far only 
explorations have been carried out to calibrate spectral types 
\citep{Bianchi2007ApJS..173..659B}.  Unfortunately, these are insufficient 
for our purposes.

\subsection{Luminosity} \label{par:lum}

From the definition of surface gravity, and eliminating the radius using the 
luminosity and effective temperature, we have 
$$\log(\rm{L/L}_\odot) = \log(\rm{M/M}_\odot) - \log g + \mbox{4}\log \rm{T}_{\rm eff} - \mbox{10.605}.$$  
High-dispersion spectroscopy gives values of $\log \rm{T}_{\rm eff}$ and $\log g$, 
but we need $\log(\rm{M/M}_\odot)$ in this formula to calculate the luminosity.  
To do this, we used the relationship expressing $\log(\rm{M/M}_\odot)$ in terms of 
$\log \rm{T}_{\rm eff}$, $\log g$ and [Fe/H] in \citet{Torres2010} (setting [Fe/H] $=$ 0.0).  
Since the values of T$_{\rm eff}$ and $\log g$ as derived from the spectra are
probably more accurate than those obtained from the Str\"omgren/Balona calibration 
and SED fitting, the $\log (\rm{L/L}_{\odot})$ values were derived from the spectroscopic 
determinations where possible.  Failing this, values determined from $uvby\beta$ 
photometry and the \citet{Balona1994} calibration were used. The adopted values of 
$\log(\rm{L/L}_\odot)$ are given in the last column of Table \ref{tab:param}.  As 
mentioned before, the spectroscopic and photometric values are in good agreement.

\begin{figure}
\centering
\includegraphics[scale=0.70]{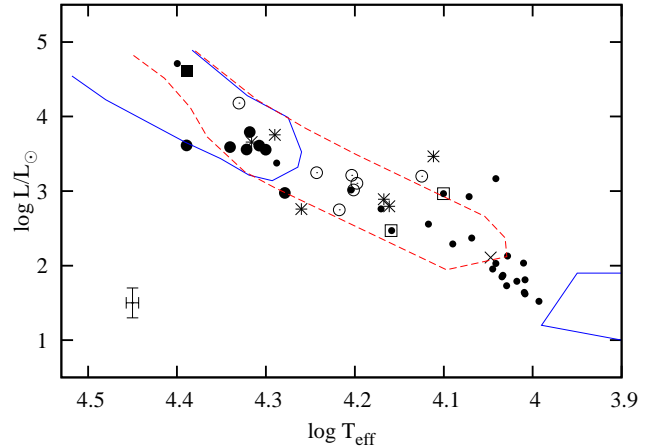}
\caption{
The theoretical $\beta$~Cep and SPB instability strips \citep[from][]{Miglio2007} 
showing the location of {\it Kepler} targets for which the stellar parameters were 
derived from ground-based follow-up observations (Table\,\ref{tab:param}).
Large filled circles are {\it Kepler} SPB/$\beta$~Cep hybrids.  Open circles
are {\it Kepler} SPB stars and the asterisks are stars with frequency
groupings.  The small filled circles are non-pulsating {\it Kepler} stars
and the cross is KIC\,11973705.  The squares with filled circles are two CP
stars.  The filled square is V1449~Aql.  The location of the empirical $\delta$~Scuti
instability strip as determined from {\it Kepler} $\delta$~Sct stars is 
indicated by the contour in the bottom right corner.  The error bars at the
bottom left-hand corner indicate a typical standard deviation of 300~K 
in effective temperature and 0.5~mag in absolute magnitude.}
\label{fig:strip}
\end{figure}

Comparison of the resulting positions in the HR-diagram with known 
$\beta$~Cep and SPB stars and with the instability strips for modes of 
low and high degree is shown in Fig.~\ref{fig:strip}.

\section{Data and frequency analysis}
\label{freqan}

Following each quarter of a 370-d {\it Kepler} solar orbit, the telescope is 
rolled to keep its solar panels facing the Sun. Choices and changes of targets are
made on the basis of these ``quarters''.  Most data are obtained in long cadence 
(LC) with integration times of 29.4\,min. For 512 targets data are obtained in 
short cadence (SC) mode with sampling times just under 1-min.  Characteristics of 
LC mode are described in \cite{Jenkins2010b} and characteristics of SC
mode in \cite{Gilliland2010}.  An overview of the {\it Kepler} science processing 
pipeline is given by \cite{Jenkins2010a}.  The duration of the quarters are as follows:\\
Q0: 10 days (2009 May 2--2009 May 11),\\ 
Q1: 33 days (2009 May 13--2009 June 15),\\ 
Q2: 89 days (2009 June 19--2009 September 16),\\ 
Q3: 89 days (2009 September 18--2009 December 16),\\ 
Q4: 90 days (2009 December 19--2009 March 19).

These observations provide a unique opportunity to study the incidence of 
pulsation at the lowest amplitude levels. The median noise level, $\sigma_{\rm M}$, 
in the periodogram depends on the stellar magnitude, $V$, and on the duration of 
the run, $\Delta t$. We find that a relation of the form: 
$$\log (\sigma_{\rm M}/\mbox{ppm}) = \mbox{0.2}(g-\mbox{8.0}) -\mbox{0.5}\log (\Delta t/\mbox{d}) + \mbox{0.9},$$ 
where $\sigma_{\rm M}$ describes the median noise level of the {\it Kepler} periodograms 
quite well.  A more practical measure of detection threshold, however, is the level 
of the top-of-the-peaks in the periodogram (the grass level), which is approximately 
equal to (3--4)$\sigma_{\rm M}$. For our sample of B stars and full Q0--Q4 coverage 
this detection threshold is typically equal to about 1~ppm for the brightest 
stars and about 15~ppm for the faintest.

\begin{figure}
\centering
\includegraphics[width=84mm]{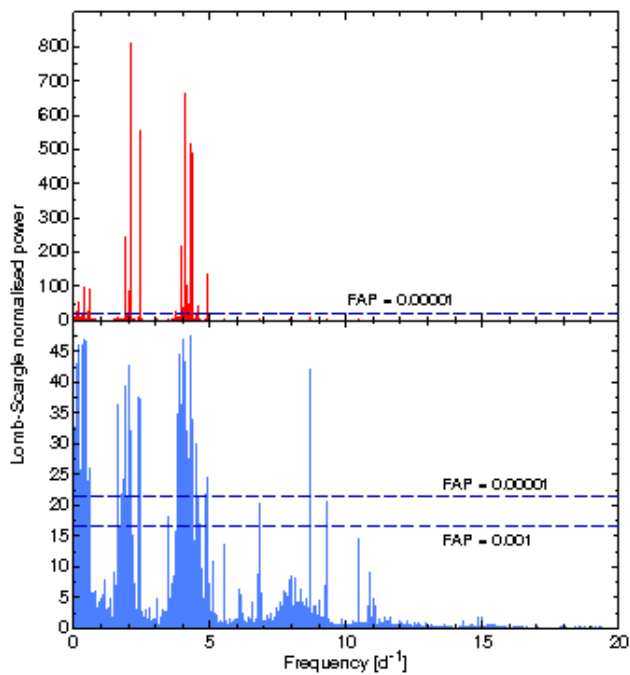}
\caption{Illustration of the determination of the false alarm probability (FAP) 
for KIC\,11360704.  The periodogram based on the Q2--Q4 data before and after 
prewhitening of 40 frequencies is shown in the top and bottom panel, respectively.
}
\label{fig:significance}
\end{figure}

In order to determine whether a peak in the periodogram is significant requires 
the calculation of a false alarm probability (FAP) which is related to the 
periodogram noise level.  A determination of the FAP, strictly valid only for 
equally-spaced data, was obtained by \citet{Scargle1982}.  Scargle's FAP requires 
the existence of a set of frequencies at which the periodogram powers are 
statistically independent. The existence of these is guaranteed only when the 
data are evenly spaced in time.  Although the {\it Kepler} data are almost 
consistently evenly spaced, the time series is interrupted for scheduled 
spacecraft rolls.  Further, observations of individual stars will contain odd 
outlier points that need to be eliminated before calculating the periodogram.
Inspection of quarters Q2--Q4 of the {\it Kepler} data (see below) show that 
approximately 5 percent of the evenly-spaced data set is lost in this way.
The conditions required for Scargle's FAP are therefore not strictly met.

Even if the Scargle conditions had been met, Scargle's method would require us 
to calculate the periodogram power {\em only} at the natural frequencies.
This restricts candidate pulsation frequencies to the Scargle grid.  The 
resolution of this grid is much poorer than that allowed by the length of the 
{\it Kepler} data set, sacrificing precision in the pulsation spectrum.
Preliminary determinations of the pulsation spectra of the stars discussed 
in this paper show a large number of very closely spaced frequencies.
The restriction imposed by the Scargle method is thus excessively debilitating 
in this case.

\citet{Frescura2008} showed that when the FAP is determined by Monte Carlo 
methods, independent frequencies are not needed, allowing the periodogram to 
be sampled on arbitrary frequency grids.  They also showed that oversampling 
of the periodogram in the frequency domain has no deleterious effect on 
frequency determination.  We computed FAPs for {\it Kepler} quarters Q2--Q4, 
using the methods detailed in \citet{Frescura2008}.  The FAPs determined by 
this method are shown in the bottom panel of Fig.\,\ref{fig:significance}, 
where they have been superimposed on a heavily prewhitened periodogram 
(with 40 frequencies) of KIC\,11360704 for illustrative purposes.
The non-prewhitened periodogram for this star is shown in the top panel 
of Fig.\,\ref{fig:significance}.  In this figure, the two lowest FAP levels 
lie so close to the frequency axis in this plot that they can not be usefully 
displayed.  The FAP levels can therefore be more clearly seen when plotted 
in Fig~\ref{fig:significance}, lower panel.  This shows the extremely high 
statistical significance of practically all the visible peaks in the top
panel of Fig.~\ref{fig:significance}, and of many dozens of peaks still 
present in the lower panel of Fig.~\ref{fig:significance}.  A detailed 
analysis of the statistical properties of the {\it Kepler} data is in 
preparation and will be published separately.

The data used here are from the short commissioning run (Q0) and the first 
four quarters (Q1--Q4).  Data for some stars are available for all these runs, 
a total observing duration of over 320~d in long cadence.  Periodograms using 
the long-cadence data have far better resolution and are more useful in 
determining the frequencies and amplitudes.  SC observations were mostly used 
to search for frequencies above 24.5~d$^{-1}$, the Nyquist frequency for the 
LC data.   We did not find frequencies higher than the long-cadence Nyquist 
frequency in any star except for KIC\,11973705.  The duration of the observing 
run for each star in short- or long-cadence mode is indicated in columns 3 and 
4 of Table \ref{tab:tar}. 

The MAST {\it Kepler} archival database provides for corrected and uncorrected 
data.  We used both corrected and uncorrected data, depending on a star and
the observing quarter.  Before the data were subjected to an analysis of the 
frequencies, the time sequence was examined and corrected for trends and 
discontinuities in the light curve.  To a large extent this is an automated 
procedure in which a smooth curve is fitted to the time series and residuals 
examined for systematic departures indicative of discontinuities.  At the same 
time, points which are clearly outliers are rejected.  While this automated 
procedure worked satisfactorily in most cases, the data corrected in this way 
were visually examined and the corrections done by hand if deemed unsatisfactory.
Owing to various instrumental effects, such as aperture changes, sudden pointing 
deviations, long-term drifts and many other effects, the data contain signals 
which do not belong to the star.  Some of the signals lead to characteristic 
peaks in the periodogram which have been tabulated by the {\it Kepler} team and 
are well-known.  Nevertheless, the behaviour at low frequencies is still poorly 
understood.  In analyzing these data, we take the view that frequencies below 
about 0.2~d$^{-1}$ should be regarded with suspicion. 

\begin{figure*}
\centering
\includegraphics{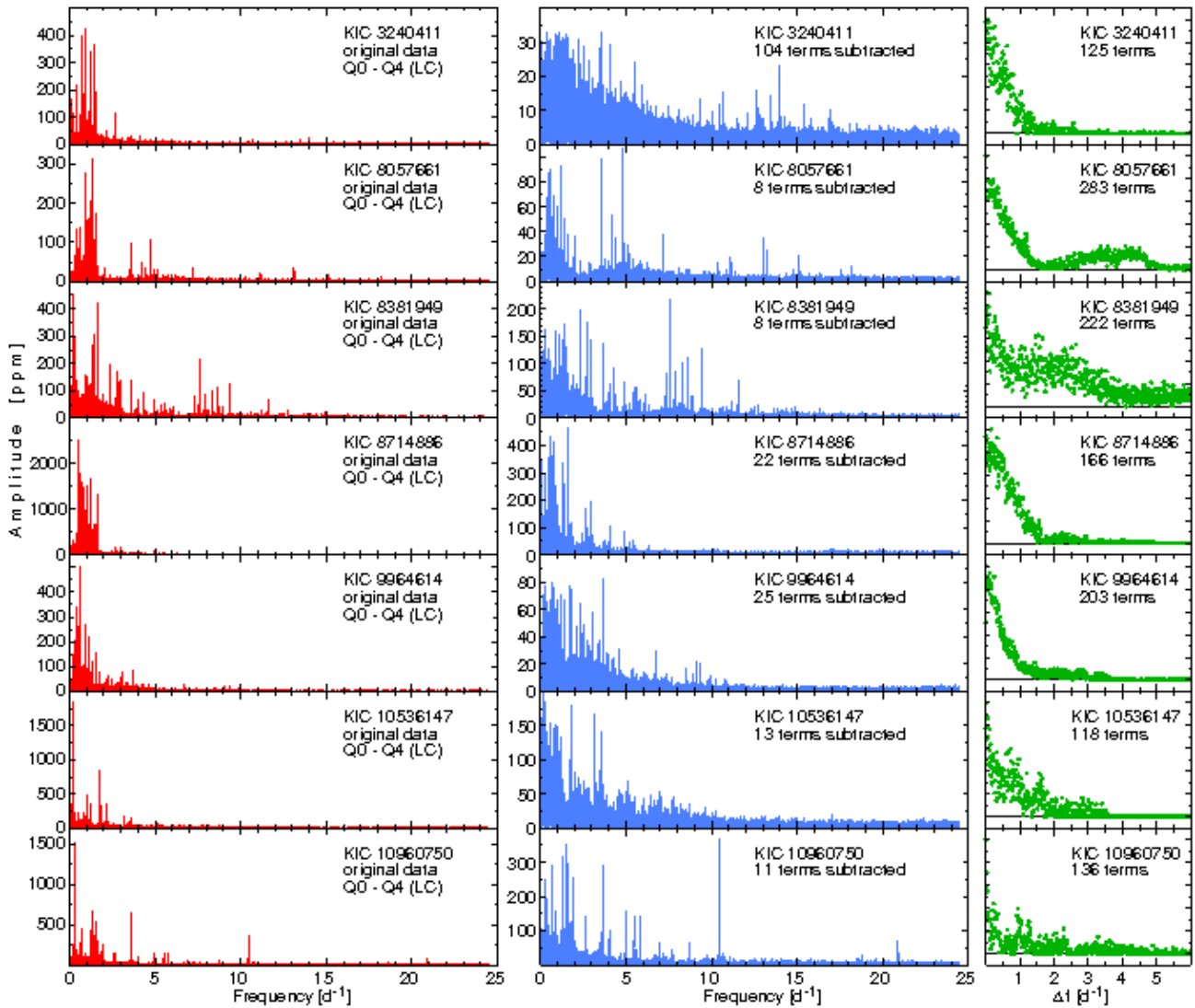}
\caption{Fourier frequency spectra of {\it Kepler} LC data for seven SPB/$\beta$~Cephei
hybrid stars.  Two periodograms are presented for each star: calculated for 
original data (left) and after prewhitening with a number (indicated) of low-frequency 
terms (middle). The periodograms were calculated up to the Nyquist frequency 
(24.5~d$^{-1}$) for the LC data. The panels on the right show histograms of weighted 
frequency spacings (HFS) plotted as a function of spacing, $\Delta f$. The numbers 
in a spacing bin are normalized to the highest bin.}
\label{fig:hybrids}
\end{figure*}

It should be noted that while these corrections are often subjective in nature, 
they only affect the low frequency range.  Frequencies above 0.5~d$^{-1}$ are hardly 
affected at all by the corrections and comparison of frequencies done independently 
by different authors are in good agreement.  One might also question the level at 
which a certain peak is deemed to be significant.  In some analyses the Lomb-Scargle 
false alarm criterion \citep{Scargle1982} was set to some level, above which peaks 
are regarded not to be significant.  In others, the local signal-to-noise level was 
calculated and the peak deemed significant only if it exceeds a certain factor.
These decisions do not affect the periodogram, but only the set of frequencies which 
are thought to be real.  In this paper, the frequencies we will be discussing are 
certainly significant by any criterion.

Characteristic frequency spacings in the periodogram, if they exist, clearly 
provide information about the star. In order to detect characteristic frequency 
spacings we calculated histograms of all possible frequency separations between 
unique frequency pairs after the frequencies were extracted.  It is necessary to 
impose some frequency and amplitude restrictions in using this method.  This was 
done by attributing arbitrary weights to the spacings. The weights depend on
the amplitude, $w_{ij} = \log (A_i \cdot A_j)$, where $A_i$ and $A_j$ denote 
amplitudes of the two terms that form a spacing and $w_{ij}$ is the weight 
attributed to the spacing  $\Delta f_{ij} = |f_i -f_j|$.  A bin of 0.01~d$^{-1}$ 
was used in these histograms which will be hereafter abbreviated as HFS.

\section{Results of the analysis}
\label{notes}
The periodograms and light curves of the stars in Table \ref{tab:tar} were
examined and deductions made which are described below.  It is, of course,
difficult to deduce the true nature of the star just from the periodogram. 
Nevertheless, with this information supplemented by the stellar parameters that
were derived for most stars under investigation, we have made an 
attempt at understanding the true nature of detected variations.

\begin{figure*}
\centering
\includegraphics{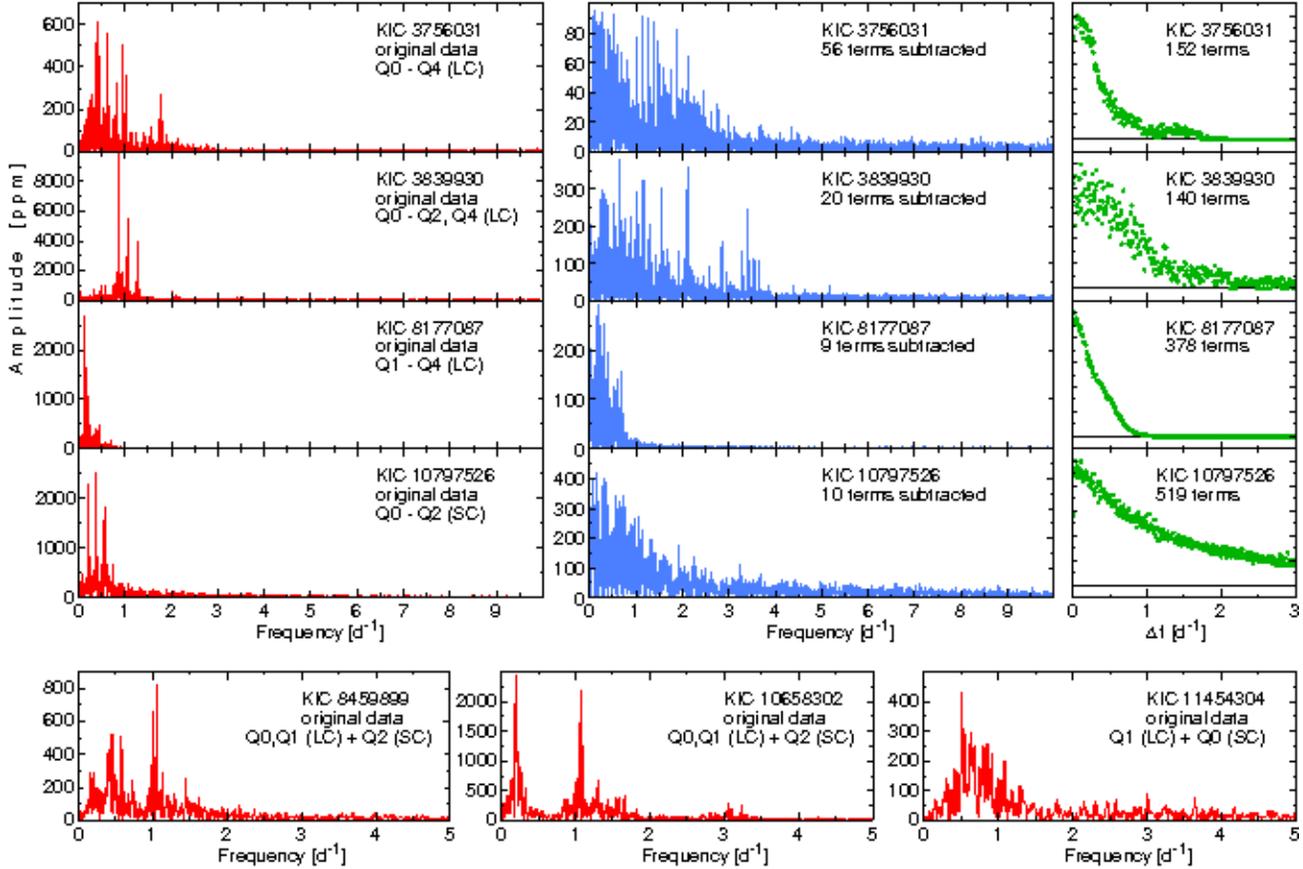}
\caption{{\it Top}: The same as in Fig.~\ref{fig:hybrids} but for SPB stars.  
{\it Bottom}: Frequency spectra for three stars with a small number of
detected terms.}
\label{fig:spb}
\end{figure*}

\subsection{SPB/$\beta$~Cephei hybrids}

When we examined the periodograms of {\it Kepler} B stars we were
immediately struck by their strange appearance.  In ground-based
observations, and even observations from {\it MOST} and {\it CoRoT}, a
$\beta$~Cep star is easily recognizable in having distinctive peaks in the
high-frequency domain.  It has long been assumed that the typical
frequencies for $\beta$~Cep stars are in the range 4--20~d$^{-1}$ with
perhaps a few low-amplitude peaks in the region 0.5--2~d$^{-1}$ which
characterize typical $\beta$~Cep/SPB hybrids.  The SPB stars, on the other
hand show distinct peaks in the low-frequency domain and no frequencies much
higher than 2 or 3~d$^{-1}$.  Here we assume that stars with frequencies
exceeding 3.5~d$^{-1}$ can be regarded as belonging to the $\beta$~Cep group
since this frequency is the approximate boundary between $p$ and $g$ modes in
non-rotating stars on the zero-age main sequence.  We realize this is an
arbitrary number, but we chose it simply for classification purposes. 

The periodograms of {\it Kepler} pulsating B stars are completely different.
They show a very large number of peaks in the low-frequency domain, but with 
some isolated low-amplitude high-frequency peaks.  Nothing like this has ever 
been seen before and we need to ask ourselves how such stars should be classified. 
They are not typical $\beta$~Cep stars because the scattered high-frequency
peaks usually have very low amplitudes and the periodogram is dominated by a
forest of low-frequency peaks.  They are not typical SPB stars either
because SPB stars do not have low-amplitude high-frequency peaks. 
They do not resemble any of the known hybrids which are really $\beta$~Cep
stars with a few isolated, weak, low-frequency peaks.  A possible solution to 
the strange appearance of the periodograms is discussed in
Section~\ref{comparison}.

There are seven stars in our sample that we will call SPB/$\beta$~Cephei hybrids
to stress that the dominant modes are the SPB-like $g$ modes (Fig.~\ref{fig:hybrids}) 
compared to the $\beta$~Cephei/SPB hybrids where the $\beta$~Cephei $p$ modes have the 
highest amplitudes.  The periodograms of the two types of hybrids are very different. 
Since low frequencies dominate in all seven {\it Kepler} stars, the middle panels of  
Fig.~\ref{fig:hybrids} show frequency spectra prewhitened with the strongest 
low-frequency modes in order to reveal the high frequency peaks more clearly. 
One assumes that the low frequencies are $g$ modes while the few isolated 
high frequencies may be normal $p$ modes. However, this assumption may not be
correct because the effect of rotation may scatter some low-frequency $g$ modes 
into the high frequency domain (as measured by the observer).  For example a 
sectorial mode of low intrinsic frequency having $l =$ 20 and azimuthal 
number $m =$ 20 may well appear at a frequency of 10--20~d$^{-1}$ since the 
rotational frequency perturbation is proportional to $m\Omega$ where $\Omega$, 
the rotational frequency, could be around 1~d$^{-1}$ in some stars.  

The detected amplitudes are very small and except for the dominant period
in KIC\,10960750 do not exceed 700~ppm.   As was indicated above, however, the 
320-day long {\it Kepler} data are of unprecedented precision in which the noise 
level is only several ppm.   The number of significant modes  that can be
extracted is therefore very large and greatly exceeds the number of
frequencies in ground-based observations of hybrid stars.
The histograms of frequency spacings are shown in the right-hand panels of 
Fig.~\ref{fig:hybrids}.  These histograms basically reflect the structure of
the periodogram in a different way and are quite interesting for tracing
similarities between one star and the next.  It is interesting to note the 
presence of a 2$f$ harmonic of the prominent high-frequency peak at 
$f=$ 10.449~d$^{-1}$ in KIC\,10960750.

Of the seven SPB/$\beta$~Cep hybrids, KIC\,3240411, 8057661, 10536147 and 10960750, 
have known values of $v\sin i$ (Table \ref{tab:param}).  The first two stars
have low projected rotational velocities typical of SPB stars, but the last
two stars have high $v\sin i$.  There are very few SPB stars known from 
ground-based observations in which $v \sin i >$ 100~km~s$^{-1}$.  In spite of 
the pronounced difference in projected rotational velocities between these 
two pairs of stars, we can find no obvious difference in either the
appearance of the periodograms or their associated frequency spacing
distributions.

KIC\,8057661 (KW\,41-38, \citealt{Kohoutek1997}), another SPB/$\beta$~Cep hybrid, 
is included in the catalogue of objects with H$\alpha$ emission \citep{Hardorp1964}.
However, recent spectra do not show any emission in the Balmer lines and the spectral 
lines are relatively narrow (J.\,Guti\'errez-Soto, in preparation). It may be 
therefore a Be star seen at low inclination which temporarily does not show 
H$\alpha$ emission.
\begin{figure*}
\centering
\includegraphics{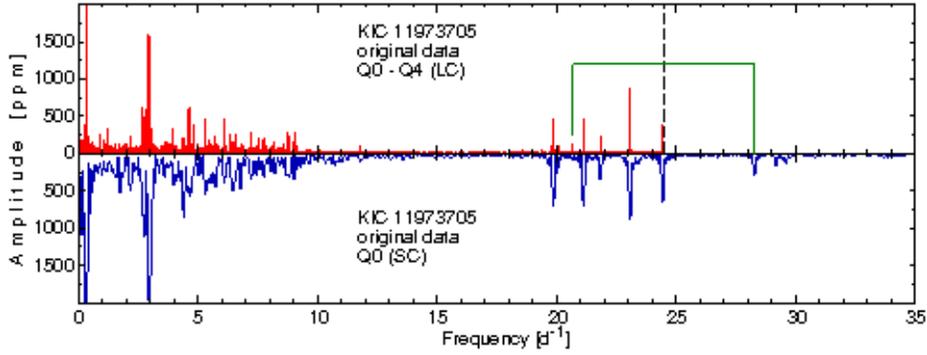}
\caption{Frequency spectra for {\it Kepler} LC (top) and SC (bottom) data of 
KIC\,11973705. The highest peak at 0.2953~d$^{-1}$ is truncated. The vertical dashed line
corresponds to frequency 24.5~d$^{-1}$, the Nyquist frequency for the LC data.}
\label{fig:kic119}
\end{figure*}

\subsection{SPB stars}
The next group of eight stars we discuss here are stars that can be called 
`normal' SPB stars because they show only low-frequencies, being devoid of the
isolated high frequencies in the stars discussed above.  Their frequency spectra 
are shown in  Fig.~\ref{fig:spb}. The three frequency spectra shown at the 
bottom of Fig.~\ref{fig:spb} are for stars where the number of detected modes 
is too small to calculate a reliable HFS.  

Only three stars from this group have known  $v\sin i$ values  which are low
in all three cases: 31~km\,s$^{-1}$ for KIC\,3756031, 22~km\,s$^{-1}$ 
for 8177087 and 53 km\,s$^{-1}$ for 8459899.  These values are
quite typical for ground-based SPB stars.

In two stars, KIC\,8177087 and 10797526, the formal solution comprises of a large 
number of low-frequency terms. These  two stars are, however, among the brightest 
in our sample. For all other bright stars ($g <$ 8.5~mag) a similar increase
of amplitude towards low frequency is observed. We therefore conclude that this 
increase of the noise towards low frequencies is due to the instrumental effects 
caused by the fact that the stars are close to the bright limit of {\it Kepler}
observations.

There are several high-frequency terms in the frequency spectrum of KIC\,3756031. 
They are barely visible in Fig.~\ref{fig:spb} because their signal-to-noise does 
not exceed 4.5 (amplitudes 8--16 ppm), though this is still significant.  The star 
is located in the middle of the SPB instability strip where $p$ mode instability 
is not expected. 

There is another star which may perhaps be regarded as an SPB
candidate, KIC\,11973705.  The frequency spectrum of the LC data (upper panel of 
Fig.~\ref{fig:kic119}) clearly shows multimode behaviour in two frequency regions: 
below 10~d$^{-1}$ and over 18~d$^{-1}$ with only some small-amplitude terms in 
between.  There are SC data available from Q0 which, however, lead  to
much poorer frequency resolution. In particular,  the group of modes around 
3~d$^{-1}$ is not well resolved in the SC data.  Nevertheless, the 
data are useful to show that high frequencies extend beyond the Nyquist frequency 
of the LC data ($f_{\rm N}=$ 24.5~d$^{-1}$). For example, the peak at 
$f_1^\prime=$ 20.68~d$^{-1}$ in the LC data (Fig.~\ref{fig:kic119}) is not
real; the true frequency of this mode is $f_1=$ 28.26~d$^{-1}$ as we see
clearly from the SC data.  The LC frequency is mirrored with respect to
$f_{\rm N}$: $f_1-f_{\rm N} = f_{\rm N} - f_1^\prime$.

The star was found to be a double-lined spectroscopic binary by \cite{Lehmann2010}. 
The variability of KIC\,11973705 is dominated by an almost sinusoidal variation 
with period $P=$ 3.3857~d and semi-amplitude of about 3560 ppm (Fig.~\ref{fig:rotbin}).  
This is very probably the orbital period or half this period.  The high frequencies 
probably belong to the A-type star (T$_{\rm eff} =$ 8\,000--9\,000~K, \citealt{Lehmann2010}) 
which is then a $\delta$~Sct variable.  One could interpret the low frequencies as 
SPB pulsations of the B-type primary, but since many $\delta$~Sct stars have low 
frequencies ($\gamma$~Dor-type variability), these could equally well belong to the 
A-type companion.  

The frequency spectrum of KIC\,10658302 
shows weak evidence for two broad peaks, at about 1.5 and 3~d$^{-1}$.
It may perhaps be classified as one of the stars with characteristic
frequency groupings discussed below.

\begin{figure*}
\centering
\includegraphics{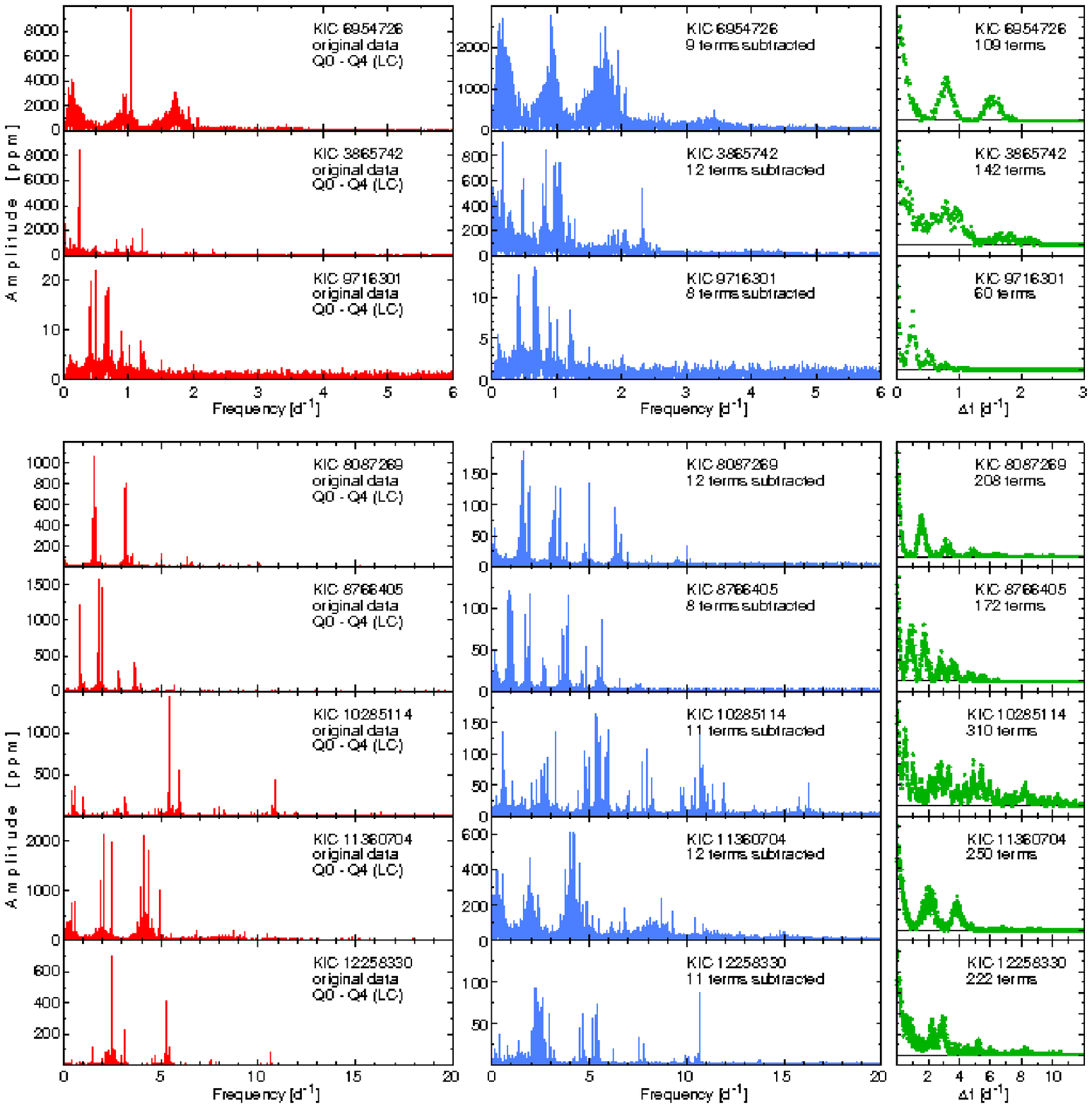}
\caption{The same as in Fig.~\ref{fig:hybrids} but for eight stars showing 
frequency groupings.}
\label{fig:group2}
\end{figure*}

\subsection{Stars with frequency groupings} \label{Be-like}

There is one certain Be star in our sample, KIC 6954726 
= StH$\alpha$~166 \citep{Stephenson1986, Downes1988, Gutierrez2010}.
Its projected rotational velocity (160~km~s$^{-1}$, Table~\ref{tab:param}) 
is moderate, so that it is probably a rather low-inclination Be star. 
Its periodogram (top panels of Fig.~\ref{fig:group2}) shows a high-amplitude 
peak at 1.027~d$^{-1}$ and three broad peaks (0.1, 0.9 and 1.7~d$^{-1}$). 
The HFS for this star shows maxima for separations $\Delta f \approx$ 0.2,
0.8 and 1.5~d$^{-1}$.  In fact, the light curve of this star is 
dominated by quasiperiodic long-term variability which was subtracted at 
the preprocessing phase. The frequency spectrum for this star seems to be
characteristic of Be stars. For instance, very similar groupings were
discovered in several Be stars observed by the {\it MOST} \citep{Walker2005,
Saio2007, Cameron2008} and {\it CoRoT} satellites \citep{Huat2009,
Diago2009, Neiner2009, Gutierrez2010}. These groupings are usually 
interpreted in terms of high radial order $g$ modes that have frequencies 
in the corotating frame smaller than the rotational frequency
\citep{Walker2005, Dziembowski2007, Cameron2008}.  Another explanation is 
that these short-term variations are due to surface or circumstellar 
inhomogeneities \citep{Balona2009} where the second and third broad peaks 
are interpreted as the rotational frequency and its first harmonic.  A 
combination of both effects is also a possibility.  In both hypotheses,
the separation of the broad peaks is interpreted as the rotational 
frequency, $\Omega$.

Seven other stars from our sample show frequency spectra which, to some 
extent, are similar to that of KIC~6954726, i.e., they show at least three 
frequency groupings as judged from their HFS (Fig.~\ref{fig:group2}).  None 
of these stars is known as a Be star.  In the HR diagram (Fig.~\ref{fig:strip}), 
they cover the range of effective temperature corresponding to B2--B8. 
Projected rotational velocities are available for four of these stars and 
are moderate to high, ranging between 133 and 271~km~s$^{-1}$
(Table~\ref{tab:param}).  For comparison, the mean projected velocities of 
B6--B9 main sequence stars ranges from 73~km~s$^{-1}$ (B9--9.5 IV) to 
144~km~s$^{-1}$ (B6--B8V) \citep{Levato2004}.  The critical rotational 
velocity for these mid- to late-B stars is about 400 km~s$^{-1}$. The mean 
equatorial velocities of Be stars is about 350~km~s$^{-1}$.  Our sample of 
these stars is too small to draw any reliable conclusion, but it would be 
indeed interesting to verify if the rotational velocities of stars in this 
group are systematically higher than the mean for B-type stars.

If either of the hypotheses proposed for the short-term variability in Be 
stars is also valid for stars with period groupings, the spacing between 
frequency groupings, $\Delta f_{\rm max}$, can be interpreted as $\Omega$. 
Thus it may be possible to estimate the rotational frequency independently of 
spectroscopic observations of $v\sin i$.  As a result, the inclination of the 
rotation axis, $i$, can be derived. In a pulsating star, knowledge of $i$ might 
help in mode identification and calculation of the visibility of the modes. 
The values of $\Delta f_{\rm max}$ from the HFS for all eight stars shown in 
Fig.~\ref{fig:group2} are presented in Table \ref{tab:rotvel}.  We note that 
for two stars, KIC\,8766405 and 12258330, two different values of $\Delta f_{\rm max}$ 
were derived. The former shows a very complicated HFS consisting of broad 
peaks with superimposed much narrower ones, the HFS for the latter has a 
double peak. 
 
We suspect that the origin of the frequency grouping in these stars could be 
rotation, though the sample at our disposal is too small to verify this idea.  
\begin{table}
\caption{$\Delta f_{\rm max}$ derived from the HFS for eight {\it Kepler} B-type targets.}
\label{tab:rotvel}
\begin{tabular}{c@{\extracolsep{3mm}}c@{\extracolsep{8mm}}c@{\extracolsep{3mm}}c}
\hline
KIC & $\Delta f_{\rm max}$  &  KIC & $\Delta f_{\rm max}$  \\
number & [d$^{-1}$] & number & [d$^{-1}$]\\
\hline
3865742 & 0.78 & 9716301 & 0.24 \\
6954726 & 0.79 & 10285114 & 0.52 and 2.76 \\ 
8087269 & 1.59 & 11360704 & 2.05 \\
8766405 & 0.88 & 12258330 & 2.24 and 2.95 \\
\hline
\end{tabular}
\end{table}

\subsection{Stars with variations due to binarity or rotation}
\begin{figure*}
\centering
\includegraphics{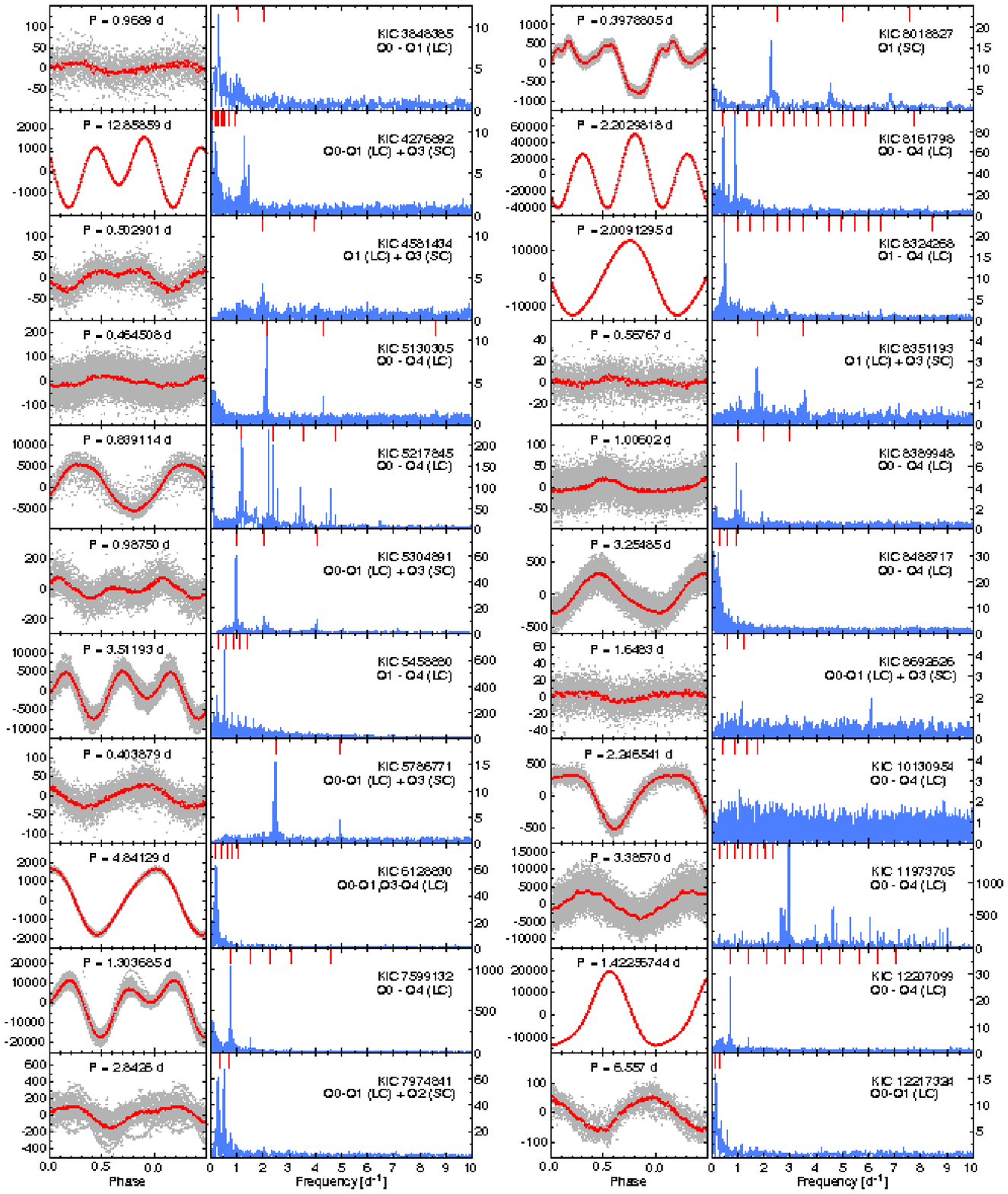}
\caption{Phase diagrams (left panels) and frequency spectra of the residuals 
(right panels) after removing the dominant periodicity shown in the left panel. 
In the phase diagrams the means in 0.01 phase intervals are also plotted.  The 
long ticks at the top of frequency spectra indicate the frequency of the principal 
periodicity and all its detected harmonics below 10~d$^{-1}$.  The ordinate in 
all panels is given in ppm.}
\label{fig:rotbin}
\end{figure*}

\begin{figure*}
\centering
\includegraphics{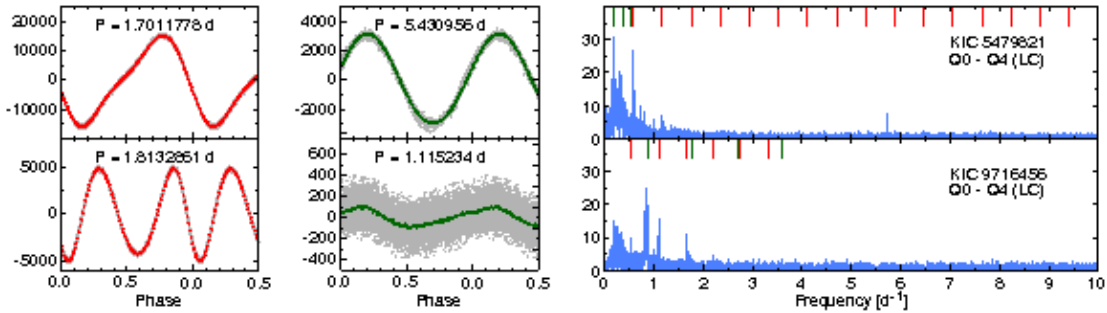}
\caption{The same as in Fig.~\ref{fig:rotbin}, but for two stars, KIC\,5479821 
and 9716456, with two non-sinusoidal periodicities detected in their light curves.}
\label{fig:rotbin2}
\end{figure*}

About a half of stars in our sample (23) show much less complicated light curves. 
Their variability can be understood in terms of proximity effects 
in a binary system, the presence of surface inhomogeneities and rotation or the 
combination of both. With the photometric precision of the {\it Kepler} data, 
one may expect to detect weak proximity effects such as light reflection, 
deformation of components and Doppler beaming \citep{Zucker2007}, even in 
low-inclination systems.   The light curves of these stars are shown in 
Figs.~\ref{fig:rotbin} and \ref{fig:rotbin2}.  The periods and amplitudes of 
the main periodicity ($A_1$) and its second harmonic ($A_2$) are given in 
Table \ref{tab:rotbin}.  Here we have included KIC\,11973705, the binary in 
which the companion is a $\delta$~Scuti variable.

\begin{table*}
\caption{List of the parameters of 24 B-type stars with variations caused 
primarily by binarity and rotation.}
\label{tab:rotbin}
\begin{tabular}{rlrrrcl}
\hline\noalign{\smallskip}
\multicolumn{1}{c}{KIC} & 
\multicolumn{1}{c}{Period} & 
\multicolumn{1}{c}{$A_1$} & 
\multicolumn{1}{c}{$A_2$} & 
\multicolumn{1}{c}{$v\sin i$} & 
\multicolumn{1}{c}{$R_{\rm min}$} & 
\multicolumn{1}{l}{Notes} 
\\
\multicolumn{1}{c}{number} &
\multicolumn{1}{c}{[d]} &
\multicolumn{1}{c}{[ppm]} &
\multicolumn{1}{c}{[ppm]} &
\multicolumn{1}{c}{[km s$^{-1}$]} &
\multicolumn{1}{c}{[R$_\odot$]} &
\multicolumn{1}{c}{}
\\
\hline
  3848385  &  0.9689(7)     &      9.5(7) & 3.1(7) & --- & --- & \\ 
  4276892  & 12.85859(6)   &  581.4(5) & 1194.6(5) & 10 & 2.53 & SB1?\\
  4581434  &   0.502901(8) &    19.2(5) & 11.6(5) & 200 & 1.98 & SB1\\
  5130305  &   0.464508(7) &    17.5(4) &   4.8(4) & 155 & 1.41 & \\
  5217845  &   0.839114(2) &5643(9) & 38(9) & 237 & 3.91 & binary\\
  5304891  &   0.98750(3)   &     34.5(14) & 37.7 (14) & 180 & 3.49 & \\
  5458880  &   3.51193(4)   &2710(16) & 4675(16) & --- & --- &\\
  5479821  & 1.7011778(4) & 13998.4(19) & 3429.5(19) & 85 & 2.84 &\\
  ...  & 5.430956(19) & 3043.2(19) & 86.1(19)& ... & 9.07 &\\
  5786771  & 0.403879(6)  & 28.1(5) & 3.6(5) & 200 & 1.59 &\\
  6128830  &  4.84129(3) & 1688.5(18) & 205.9(18)& 15 & 1.43 &CP\\
  7599132  &  1.303685(3) & 8890(20) & 8458(20) & 63/50 & 1.61/1.28 &\\
  7974841  &  2.8426(5) & 107.9(26) & 41.1(26) & --- & ---& \\
  8018827  &  0.3978805(9) & 437.3(4) & 415.2(4) & --- & ---&\\
  8161798  &  2.2029818(4) &13208(4) & 39551(4) & --- & ---& \\
  8324268  &  2.0091295(4) &13185.7(7) & 990.0(7) & --- & ---& CP\\
  8351193  &  0.56767(4) & 1.90(24) & 2.10(24) & --- & ---&\\
  8389948  &  1.00602(3) & 12.98(26) & 4.50(26) & 142/130 & 2.81/2.57 &\\
  8488717  &  3.25485(7) & 287.7(12) & 45.2(12) & --- & ---&\\
  8692626  &  1.6483(3) & 4.86(27) & 1.23(27) & --- & ---& \\
  9716456  &  1.8132861(7) & 859.0(11) & 4455.4(11) & --- & ---&\\
  ... &  1.115234(25) & 84.6(11) & 17.3(11)& --- & ---&\\
 10130954 &  2.246541(10) & 406.7(5) & 106.5(5) & --- & --- &\\
 11973705 &  3.38570(8) & 3558(14) & 33(14) & 103 & 6.85 & SB2\\
 12207099 &  1.42256744(14) & 16159.4(7) & 3283(7)& 43 & 1.20 & SB2?\\
 12217324 &  6.557(7) & 53.3(8) & 4.1(8) & --- & ---&\\
 \hline
\end{tabular}
\end{table*}

None of the light curves of stars in this group shows evidence of eclipses. A 
thorough discussion and analysis of the binary light curves is beyond the
scope of this paper and would probably require time-series radial velocity 
observations.  Nonetheless, four stars from this group (KIC\,4276892, 4581434, 
5217845 and 12207099) are known (or suspected) to be binaries (Table \ref{tab:tar}). 
Without corresponding radial velocity observations we cannot be sure that
the photometric period corresponds to the orbital period.  We also suspect 
that other stars showing very regular variability with relatively large 
amplitudes, especially those that have double-wave light variations, are 
probably binaries as well. These include KIC\,5458880, 7599132 and 8161798 
(Fig.~\ref{fig:rotbin}) and also KIC~5479821 and 9716456
(Fig.~\ref{fig:rotbin2}).

The longest period detected in our sample is 12.9~d in KIC\,4276892, a suspected 
SB1 binary \citep{Catanzaro2010}.   Binaries with longer orbital periods may
have a photometric signature in the {\it Kepler} data, but if the amplitude
is too small it will probably be mistaken for an instrumental trend.

There are several variables (KIC\,4581434, 5130305, 5786771, 8018827, 8351193) 
with periods in the range of 0.4--0.6~d.   Among these, only KIC\,4581434 
is known to be a binary \citep{Catanzaro2010}. All but KIC\,8018827 show variability 
with the peak-to-peak amplitude of less than 30~ppm. This very small
amplitude may be due star spots or other inhomogeneities in a rotating star.
If the star spots migrate with respect to each other this could explain
amplitude  modulation in the light curves.   Amplitude modulation is present in 
the five stars mentioned above.  Their frequency spectra (Fig.~\ref{fig:rotbin}) 
show residual power in the vicinity of the subtracted main periodicity and its 
harmonics.  KIC\,8018827 is exceptional because the strongest residual signal is 
observed at a frequency of 2.269~d$^{-1}$,  i.e., about 10 percent less than 
the dominant frequency.

Similar variability can also be seen in stars that have longer periods, e.g., 
KIC\,5304891 and 8389948.  For stars where $v\sin i$ is known, an additional 
argument in favour of rotation as the cause of variability is the fact that 
the observed periods are consistent with $v\sin i$.  With the assumption that 
the main periodicity corresponds to the rotation period, the minimum radius can be 
estimated: $R_{\rm min} = P_{\rm rot} v\sin i$/2$\pi$. The values of 
$R_{\rm min}$ are given in Table \ref{tab:rotbin}  for all stars that have 
measured $v\sin i$.  We note that the radii of late B-type stars on 
the main sequence range from about 1.5~R$_\odot$ to 3.5~R$_\odot$. 
On this basis, rotation as the cause of variability seems to be excluded for
KIC\,11973705, the longer period of KIC\,5479821 and also perhaps for KIC\,5217845. 
The first star and the last star are known binaries, so we suspect that their 
variability originates from this cause.

KIC\,5479821 is one of two stars (Fig.~\ref{fig:rotbin2}) which shows two 
non-sinusoidal periodicities. If the longest period in these two stars is 
attributed to orbital effects, the shortest period needs to be interpreted 
differently.  Rotation is one possibility, but if this is the case, the systems 
are strongly non-synchronously rotating.  Contamination or multiplicity cannot 
be excluded either. Variability caused by rotation and binarity are not mutually 
exclusive.  In short-period binaries, synchronization  of the orbital and 
rotation period can be expected, but because B-type stars are relatively young, 
some deviations from full synchronization is possible.  In this context, we 
note that KIC\,5217845, which is a binary \citep{Lehmann2010}, has a light 
curve which is strongly modulated with a period of about 6 days.

Among the stars discussed in this section there are two chemically peculiar (CP) 
stars.  KIC\,8324268 (HD\,189160, V2095~Cyg), was discovered as a CP Si star by 
\cite{Zirin1951}. Its photometric variability was detected in {\it Hipparcos}
data, but the period could not be derived.  In {\it Kepler} data the star shows 
strictly periodic variability with a period of almost exactly 2 days 
(Fig.~\ref{fig:rotbin}, Table \ref{tab:rotbin}) and only very small
amplitude modulation of about 0.2 percent. The period is assumed to
be the rotational period as in other $\alpha^2$~CVn stars.

The other CP star is KIC\,6128830, a HgMn star \citep{Catanzaro2010}. 
The light curve is nearly sinusoidal with a period of about 4.84~d and peak-to-peak 
amplitude of 3600~ppm.  The photometric period, if assumed to be the
rotational period, is consistent with $v \sin i =$ 15~km\,s$^{-1}$ (Table
\ref{tab:rotbin}), which is typical for the slowly rotating HgMn stars 
\citep{Abt1972}. Light variations in HgMn stars have only recently been 
detected in {\it CoRoT} observations \citep{Alecian2009} and line profile 
variations are known only in very few stars \citep{Hubrig2008, Briquet2010}.  
The light variation implies an inhomogeneous distribution of various elements 
in the atmospheres of HgMn stars which challenges our understanding of the 
nature of these stars.  HgMn stars were known not to have strong magnetic 
fields \citep{Auriere2010} and, while diffusion should lead to vertical 
stratification, no horizontal surface abundance patches are expected.  
However, recent observations \citep{Hubrig2010} indicate the presence of 
a weak longitudinal magnetic field in the HgMn star AR~Aur.

As shown in Fig.~\ref{fig:strip}, stars from this group are mostly late 
B-type stars falling outside both the $\beta$~Cephei and SPB instability 
strips.  This may explain the lack of variability other than that related 
to rotation or binarity in these stars.  Those that are located within the 
instability strips are either CP stars or do not show variations that can be 
interpreted as due to pulsations.  Only KIC\,5479821, which is located in the 
SPB instability strip, shows a single peak at 5.709~d$^{-1}$ and amplitude of 
8~ppm which is difficult to understand.  The variability in KIC\,11973705 has 
already been discussed.

On the other hand, there are frequencies that cannot be easily explained in 
four late B-type stars:  KIC\,3848385 (0.31~d$^{-1}$), 4276892 (1.29 and 
1.45~d$^{-1}$), 7974841 (0.28 and 0.52~d$^{-1}$), and 8692626 (6.09~d$^{-1}$).  

Finally, we are left with two stars, KIC\,9655433 and 11817929, which we have 
not yet discussed. The former is the only one from our sample which shows no 
evidence of variability in {\it Kepler} data. The latter is a member of the open 
cluster NGC\,6811 and shows the frequency spectrum characteristic of $\delta$~Sct 
stars.  Its spectral type is very poorly known (Table \ref{tab:tar}) and we 
suspect that it may be an A-type star.

\section{Modes of high degree in B-type stars}
\label{comparison}

We have seen that there are several B stars in the {\it Kepler} data base which 
are clearly SPB or SPB/$\beta$~Cep hybrids according to our definition.  We
have already remarked on the vastly different appearance of the periodograms
which are so different from those in ground-based observations of these
stars.  This requires an explanation.

We see from Fig.~\ref{fig:strip} that all of the SPB/$\beta$~Cep hybrid stars 
are within the $\beta$~Cep instability strip, with KIC\, 8714886 just cooler 
than the red edge, but within the SPB instability strip.  The instability strips 
generally discussed in the context of ground-based observations are constructed 
for modes with $l \le$ 3.  From the ground we only see modes of high amplitude, 
and these tend to be modes with $l \le$ 2 because cancellation effects render 
modes with higher $l$ invisible in photometry.  The amplitude limit for ground-based 
observations is typically a few millimagnitudes, but for {\it Kepler} photometry 
the detection limit is only a few ppm.  At this level, the visible range of 
$l$ must be considerably increased.  Cancellation effects are severe up to 
$l \approx$ 4, but for higher values of $l$ the decrease in visibility 
with $l$ is much smaller \citep{Balona1999}.  It is possible that with 
{\it Kepler} data, pulsations with values of $l$ as high as 20, or even 
higher, may be visible.    Why a particular mode should be selected in
preference to another is, at present, an unsolved problem and there is no
reason to suppose that only modes of low degree are present in these stars.
Clearly, we cannot limit the comparison with models to low values of $l$.

One of the puzzling aspects of the {\it Kepler} SPB/$\beta$~Cep hybrids is 
the fact that their frequency spectra are dominated by low frequencies.
On the other hand, ground-based observations of SPB/$\beta$~Cep hybrids are 
dominated by the high frequencies with only a few low-frequency SPB-type modes.
\cite{Balona1999} have studied the excitation of modes of high $l$ in B-type 
stars.  They find that for a given value of $l$ there are two separate frequency 
ranges where unstable modes may occur: a range of SPB-like low frequencies and a 
high-frequency ($\beta$~Cep like) range.  In the high-frequency range only modes 
of low $l$ are excited, while in the low-frequency range modes of high $l$ 
are excited.  These high-degree modes are trapped in the outer layers.

In zero-age main sequence stars the region of SPB-type low-frequency instability 
is present for stars with 4.1 $< \log (\rm{T}_{\rm eff}/\mbox{K}) <$ 4.4 (B7 to B1) 
and reaches its maximum extent for $l =$ 8.  As the star evolves, the SPB-type 
region extents further to hotter temperatures up to $\log \rm{T}_{\rm eff} =$ 4.5 
(O9, B0), reaching its maximum extent for $l =$ 10.  The low-frequency region is 
still present, though smaller, for $l =$ 24.  It turns out that the maximum extent 
of the instability strip for low-frequency $g$ modes for $l$ up to 24 more or 
less coincides with the SPB instability strip for low $l$ shown in Fig.\,\ref{fig:strip}.   
It therefore covers almost the entire B star region.  Thus it is not surprising 
to find that most of the stars observed by {\it Kepler} show many low-frequency 
modes.  In all likelihood, these are $g$ modes with high $l$.  They are
not seen from the ground because of their low amplitudes.

It would seem that none of the stars observed by {\it Kepler} are in the region 
of the HR diagram where $p$ modes are predominantly excited.  The hottest stars 
in our sample all lie on the cool side of the $p$-mode $\beta$~Cep instability 
strip as can be seen in Fig.\,\ref{fig:strip}.  The high-frequency modes are
perhaps only weakly driven and have very low amplitudes.  Of course, rotation 
also plays a big role, especially for modes of high $l$ which can be perturbed 
to higher and lower frequencies in proportion to the rotational frequency.
The peculiar appearance of the periodograms of these stars thus has a logical
explanation in terms of excitation of high-degree modes and their location
on the cool side of the $\beta$~Cep instability strip.

In Fig.\,\ref{fig:strip} we also show the location of {\it Kepler} B stars which 
do not pulsate.  As already mentioned, some of these stars may be binaries, but 
there is nothing that clearly distinguishes pulsating and non-pulsating stars.
In ground-based observations, the SPB stars nearly all have $v \sin i <$ 100~km~s$^{-1}$,
which is not a selection effect and suggests that high rotation suppresses pulsation.
The {\it Kepler} sample of pulsating and non-pulsating stars comprise both 
slow and rapid rotators, so rotation does not appear to be a factor.  We
note that some of these non-pulsating B stars are within the SPB strip and
two are within or close to the $\beta$~Cep instability strip.  This
indicates that the instability strips are not pure, i.e. that stars need not
pulsate even if they are located within the instability strip.

\section{Stochastic excitation of modes}
\label{stochastic}

As we mentioned in the introduction, one of the most interesting results from 
space observations of $\beta$~Cep stars was the possible discovery of completely 
unexpected modes with short lifetimes in {\it CoRoT} observations of the 
B1.5\,II-III $\beta$~Cep star V1449\,Aql \citep{Belkacem2009} and in the 
O-type star HD\,46149 \citep{Degroote2010A&A...519A..38D}. Features in the 
periodograms of V1449\,Aql were found in the range 9--22~d$^{-1}$ and were 
interpreted as stochastic modes excited by turbulent convection just below the 
photosphere in the thin Fe and He convective zones 
\citep{Cantiello2009,Belkacem2010}. The occurrence of convection 
in the region of iron opacity bump depends strongly on the luminosity and 
metallicity of the star \citep{Cantiello2009}. One thus expects that the 
occurrence of stochastic oscillations in B stars would follow a similar 
dependence if the modes are excited in the Fe-bump region. 

Confirmation of the existence of stochastically excited modes in massive 
stars is clearly of great interest. One of the problems facing such an 
investigation is that it is not a simple matter to distinguish between 
self-excited and stochastic modes, as well as non-pulsational light variations. 
If oscillations occurred in the same frequency range as the $\beta$~Cep pulsations, 
as they do in V1449\,Aql, then one cannot immediately eliminate the Fe bump 
$\kappa$~mechanism as the excitation mechanism for these modes. The suspected 
stochastic modes seen in the {\it CoRoT} observations appear as wide structures 
in the periodograms of V1449\,Aql and HD\,46149. They can either be interpreted 
as a large number of self-excited modes driven by the $\kappa$~mechanism 
(possibly beating with a rotation period unresolved in the measurements) 
or as single modes that change amplitude and phase stochastically. It is not 
possible to distinguish between these two explanations unless further 
information is available.

In \cite{Belkacem2009} the evidence is presented as a time-frequency diagram in 
which the constant amplitude of a nearby self-excited mode is contrasted with the 
variable amplitude of the modes thought to be stochastically excited.  These 
authors find that the stochastic modes are organized as regularly spaced patterns 
in the periodogram, which is stated to be signature of these modes.  Regular 
spacing applies only in the asymptotic limit of high frequencies and occurs for 
both stochastic and self-excited modes.  The claimed modes in V1449\,Aql may well 
be stochastically excited, but unfortunately it cannot be proven at this stage. 
What lends credence to such an hypothesis, however, is that there is a prediction 
of where stochastic modes are expected to occur in the HR diagram among the B 
stars \citep{Cantiello2009}.  It would certainly be important to search for wide 
structures in the periodograms of the {\it Kepler} B stars to determine if such 
stars conform to this prediction.

We visually examined the periodogram of {\it CoRoT} observations of V1449\,Aql to 
familiarize ourselves with the broad structures.  We then looked for similar 
structures in the periodograms of the pulsating stars shown in Fig.~\ref{fig:hybrids}, 
but nothing resembling them could be found.  All the stars have well-resolved peaks 
in the region 10--24~d$^{-1}$.  Moreover, the background noise level conforms to 
the relationship discussed in Section~\ref{freqan} except for KIC\,3865742 and 
KIC\,10960750 where it is about a factor of 3 or 4 higher than expected for the 
stellar magnitude and length of the time series.  For both stars this is a result 
of the huge number of low frequencies.  There are at least 600 signals with 
frequencies less than 6~d$^{-1}$ in the {\it Kepler} light curves of these two stars.  

KIC\,5458880 has an effective temperature and luminosity very similar to
V1449\,Aql (T$_{\rm eff} =$ 24\,500\,K, $\log g =$ 3.45,  \citealt{Morel2007}).  
It has a strange double-wave light curve which we have interpreted as a proximity 
effect of a binary.  The periodogram shows a noise level which increases towards 
low frequencies with no clear evidence of distinct frequencies, except at the 
binary period and its harmonic, down to a level of about 10~ppm.  There is 
certainly no sign of broad structures.  The peaks we examined have amplitudes 
of about 10~ppm, i.e., an order of magnitude smaller than those found in V1449\,Aql.

Our conclusion is that the existence of stochastic modes in B stars is not yet 
proven and that the {\it Kepler} stars show no indication of the presence of these 
modes.   If the modes do exist, this result may simply be due to the fact that the 
{\it Kepler} stars in which $\beta$~Cep pulsations are seen are too cool to have a 
sufficiently active convection zone.  The occurrence of stochastic oscillations 
relies on the occurrence of convection in the iron opacity bump region. 
Since we cannot yet explain the occurrence of constant stars within the
$\beta$~Cep instability strip, it is also possible that a star may not
pulsate even if it is in a location in the HR diagram where stochastic
driving is expected to be efficient.

\section{Conclusion}
\label{conclusion}

{\it Kepler} observations of 48 main sequence B stars show that 15 of them 
pulsate with frequencies typical of SPB stars. Of these stars, seven show
isolated, weak, high frequencies in the $\beta$~Cep range. The frequency spectra 
differ considerably from those seen in ground-based SPB/$\beta$~Cep hybrids
and can be understood if most of the low-frequency modes are of high
spherical harmonic degree.  Models show that low-frequency modes of high 
spherical harmonic degree, $l$, up to $l =$ 24 are excited 
in stars spanning practically the whole main sequence band in B stars. 
Since we expect to see far more modes of high $l$ in {\it Kepler} observations 
than in ground-based observations, this may explain why so many low-frequency 
modes are visible in the {\it Kepler} data.  Although the presence of low- and 
high-frequency modes place these stars in the same class as the ground-based 
SPB/$\beta$~Cep hybrids, it is not clear whether this classification is 
meaningful in the context of space observations.  We call the hybrids
observed by {\it Kepler} SPB/$\beta$~Cep hybrids  to emphasize the dominance
of modes in the SPB region.

We also find that though nearly all 48 B stars in our sample are variable at 
some level, there are certainly stars within the SPB as well as the cooler side 
of the $\beta$~Cep instability strip which are not SPB stars by the usual 
definition of these stars. This indicates that there are mechanisms which 
restrict pulsations in B stars. Such mechanisms may be rotation (which appears 
to inhibit pulsations in SPB stars), magnetic fields and chemical stratification.  
It would be important to determine which (if any) of these factors are responsible 
for damping of pulsations by a comparative study of pulsating and non-pulsating 
stars.  We may also conclude from these observations that the $\beta$~Cep instability 
strip is not pure, i.e., there are stars in the $\beta$~Cep instability strip 
that do not pulsate at all (KIC\,5458880 and 10130954).  This conclusion is 
strengthened by the B0.5\,IV star HD\,51756 for which the variations observed in
the {\it CoRoT} light curve are interpreted as due to rotational modulation 
(Pap\'ics et al., submitted).  There are also many stars within the SPB 
instability strip which do not seem to pulsate. 

We have not found any indication of pulsating star situated between the hot 
end of the $\delta$~Sct instability strip and the cool end of the SPB instability 
strip. This conclusion is not only a result of the study of B stars reported here, 
but also follows from a study of main sequence A stars in the {\it Kepler} field 
not reported here. \cite{Degroote2009} found a number of such stars.  One star 
that show clear high-amplitude $\delta$~Sct pulsations, KIC\,11973705, is 
almost certainly a B star with a $\delta$~Sct companion.

We have identified a group of stars which show a structure in the periodogram 
consisting of unresolved frequency groupings which may or may not be due to 
pulsation.  The light curve shows a period corresponding to the mean frequency of 
the first broad peak (not counting the lowest frequencies) with a strongly modulated 
amplitude.  The structure of the periodogram in many cases resembles that of Be 
stars, but those stars with spectra show that they are not Be stars. Moreover, 
a similar structure is found among {\it Kepler} A and F stars as well.  The HFS 
is a valuable tool in studying  such structures and may possibly be applied in 
other cases as well.  However, the physical meaning of these structures is not 
presently understood, and without further study it does not seem worthwhile to 
pursue this line of investigation.

We have looked for the broad structures seen in {\it CoRoT} observations of 
the $\beta$~Cep star V1449~Aql and the young O-type binary HD\,46149. 
They were interpreted as solar-like stochastic modes by \citet{Belkacem2009} 
and \citet{Belkacem2010}, respectively.  We are unable to find such structures 
in the {\it Kepler} stars, but this may be because they are too cool. 
A problem is that it is not really possible to distinguish between 
quasi-periodic (non-pulsational) variations, stochastic oscillations and 
closely-spaced unresolved self-excited oscillations.  A conclusive demonstration 
that stochastic modes are present in B stars is not possible in principle
with existing data.  However, it is possible to test the hypothesis that 
convectively-driven stochastic modes occur within a certain region in the 
HR diagram by comparing the location of stars with broad features in the 
HR diagram to the predictions. If the comparison supports the hypothesis, 
then the concept of stochastic pulsations in B stars would certainly become 
more acceptable.  At this stage, such a comparison is not possible with 
{\it Kepler} data. 

It will be very difficult to use the observed frequencies on their own for
asteroseismological modelling without mode identification for at least some
frequencies.  Because of the low amplitudes, mode identification requires
multicolour space observations which will not be available in the
foreseeable future.  Some progress might be possible if systems of equally
spaced frequencies or periods can be found.  A preliminary investigation
suggests that this might be possible in one or two stars.  If this can be
confirmed, it might tell us something about the rotational splitting or
perhaps even lead to an estimate of the large separation.

Finally, we have detected variability in many stars which we suspect may be
proximity effects due to binarity.  In other stars, the modulated light
curves are more easily explained in terms of migrating star spots. We have
also detected periodic light variations in two chemically peculiar stars.
{\it Kepler} observations of B-type stars have produced some surprises as
well as new opportunities for a better understanding of these stars.

\section*{Acknowledgments} 

The authors wish to thank the {\it Kepler} team for their generosity in allowing 
the data to be released to the {\it Kepler} Asteroseismic Science Consortium (KASC) 
ahead of public release and for their outstanding efforts which have made these 
results possible.  Funding for the {\it Kepler} mission is provided by NASA's Science 
Mission Directorate.  We particularly thank Ron Gilliland for his tireless work on 
behalf of KASC.

LAB wishes to thank the South African Astronomical Observatory for financial support. 
AP acknowledges the financial support from the MNiSzW grant N\,N203 302635. 
MB is a Postdoctoral Fellow of the Fund for Scientific Research of Flanders (FWO), 
Belgium.  JN and RSz acknowledge the support by the National Office for Research 
and Technology through the Hungarian Space Office Grant No.~URK09350 and the 
`Lend\"ulet' programme of the Hungarian Academy of Sciences.

The GALEX data presented in this paper were obtained from the Multimission 
Archive at the Space Telescope Science Institute (MAST). STScI is operated by the 
Association of Universities for Research in Astronomy, Inc., under NASA contract 
NAS5-26555.  Support for MAST for non-HST data is provided by the NASA Office of 
Space Science via grant NNX09AF08G and by other grants and contracts.

\bibliographystyle{mn2e}
\bibliography{bstars} 
\label{lastpage}
\end{document}